\documentclass[
 reprint,
 superscriptaddress,
 nofootinbib,
 amsmath,
 amssymb,
 aps,
 prd,
]{revtex4-2}

\usepackage{graphics,epsfig,psfrag}
\usepackage{color}
\usepackage{graphicx}
\usepackage{dcolumn}
\usepackage{bm}
\usepackage{hyperref}
\usepackage{xcolor}
\usepackage{comment}
\usepackage{enumitem,soul}

\hypersetup{
           breaklinks=true,   
           colorlinks=false,   
           pdfusetitle=true,  
        }

\begin{document}

\title{Light deflection in axially symmetric stationary spacetimes\\
filled with a moving medium}

\author{Christian Pfeifer}
\email{christian.pfeifer@zarm.uni-bremen.de}
\affiliation{ZARM, University of Bremen, 28359 Bremen, Germany}

\author{Barbora Bezd\v{e}kov\'{a}}
\email{bbezdeko@campus.haifa.ac.il}
\affiliation{Department of Physics, Faculty of Natural Sciences, University of Haifa, Haifa 3498838, Israel}
\affiliation{Haifa Research Center for Theoretical Physics and Astrophysics, University of Haifa, Haifa 3498838, Israel}
\affiliation{Institute of Theoretical Physics, Faculty of Mathematics and Physics, Charles University, 18000 Prague, Czech Republic}

\author{Oleg Yu.~Tsupko}
\email{tsupkooleg@gmail.com}
\affiliation{ZARM, University of Bremen, 28359 Bremen, Germany}

\date{\today}

\begin{abstract}
The deflection of light rays near gravitating objects can be influenced not only by gravity itself but also by the surrounding medium. Analytical studies of such effects are possible within the geometrical optics approximation, where the medium introduces additional light bending due to refraction. These studies typically assume a cold non-magnetized plasma, for which light propagation is independent of the medium’s velocity. In this paper, we extend the analysis to the general case of dispersive refractive media in motion and study its influence on light deflection. We consider an axially symmetric stationary spacetime filled with a moving medium, motivated by the interplay between rotational effects originating from the spacetime and those induced by the medium's motion. We begin by analyzing light deflection in the equatorial plane of a rotating object in the presence of a radially moving and rotating medium. Assuming a specific form of the refractive index enables a fully analytic treatment. In the particular cases of either pure radial or pure rotational motion, we obtain explicit expressions for the deflection angle. Next, we analyze the case of a slowly moving medium and identify two particularly interesting results. 
First, we show that, to the first order in the medium's velocity, the radial motion does not affect the light deflection. Second, assuming slow rotation of the gravitating object, we demonstrate that the black hole rotation and the medium motion can produce equivalent observational signatures. We find the quantitative condition under which these effects compensate each other. This relation becomes particularly clear for a Kerr black hole, discussed as an example.
\end{abstract}

\maketitle

\section{Introduction}\label{sec:intro}

The bending of light rays by gravity is a fundamental prediction of the general theory of relativity and has been confirmed through many types of observations. Gravitational light deflection underlies a wide range of astrophysical phenomena, collectively studied under the field of ``gravitational lensing'' \cite{GL1, Blandford-Narayan-1992-review, Wambsganss-1998-review, GL2, Bartelmann-2010-review, Dodelson-GL, Congdon-Keeton-book-2018, Meneghetti-2021}. In this context, a ``lens'' refers to a massive object whose gravitational field bends the trajectory of light passing nearby. In the weak-deflection regime, lenses can range from the Sun and other stars to galaxies and galaxy clusters. Additionally, ultracompact objects such as black holes allow for the strong-gravity regime; they bend light so strongly that new lensing phenomena emerge. These include the black hole shadow \cite{Falcke-2000, Cunha-Herdeiro-2018, Perlick-Tsupko-2022, Vagnozzi-2023-review} and the so-called higher-order images \cite{Darwin-1959, Atkinson-1965, Luminet-1979, Ohanian-1987, Virbhadra-2000, Bozza-2002, Bozza-2010-review, Johnson-2020, Gralla-Lupsasca-2020a-lensing, Gralla-Lupsasca-Marrone-2020, Hadar-2021-photon-rings, BK-Tsupko-2022, Tsupko-2022-shape, Kocherlakota-2024a, Kocherlakota-2024b, Aratore-2024}.

During the propagation of light through the universe, from a source to an observer, it can be affected by various factors beyond the gravitational field of the objects it passes by. In an astrophysical context, additional influence may arise from the presence of a plasma medium through which the light propagates. If one limits the analysis to the framework of geometrical optics, plasma affects light primarily through refraction, leading to additional bending of the rays. Moreover, since plasma is a dispersive medium, all related effects become chromatic, meaning they depend on the wavelength of the light. Besides the astrophysical situation, such chromatic effects are also studied within quantum gravity phenomenology, where the propagation of light on quantum spacetime is described through a modified dispersion relation \cite{Addazi:2021xuf, Glicenstein:2019rzj, Pfeifer:2018pty, Amelino-Camelia:2023srg, Barcaroli:2017gvg, Laanemets-Hohmann-Pfeifer-2022}, that introduces an effective medium to describe the interaction between light and quantum features of gravity. 
The difference between the plasma and the quantum spacetime is the wavelength for which the effect is expected to become relevant. Searches for light-propagation effects from quantum gravity are usually carried out in the gamma-ray regime \cite{Bolmont:2022yad,Caroff:2024zvr}.

Studies of gravitational lensing in plasma consider scenarios where light rays are emitted from a source, travel towards an observer, and are affected along the way by both the gravitational lens and the surrounding plasma. In these studies, the plasma is treated as a transparent medium with a specific refractive index, while the mass of plasma particles and interaction effects, such as absorption and scattering of photons, are neglected.

Gravitational lensing in plasma has become a popular research topic in recent years, leading to a growing body of literature on the subject. Various gravitational objects and plasma distributions have been studied. In the following section, we recall previous studies which investigated axially symmetric spacetimes in the presence of cold plasma and related extension. For examples of other recent works, we refer to \cite{BK-Tsupko-2009, BK-Tsupko-2010, Tsupko-BK-2013, Morozova-2013, Er-Mao-2014, Rogers-2015, Rogers-2017a, Rogers-2017b, Schulze-Koops-Perlick-Schwarz-2017, Crisnejo-Gallo-2018, Crisnejo-Gallo-Rogers-2019, Crisnejo-Gallo-Villanueva-2019, Crisnejo-Gallo-Jusufi-2019, Kimpson-2019a, Kimpson-2019b, Sareny-2019, Turyshev-2019a, Turyshev-2019b, Wagner-Er-2020, Tsupko-BK-2020-microlensing, Er-Yang-Rogers-2020, Matsuno-2021, Er-Mao-2022, Guerrieri-Novello-2022, Chainakun-2022, Crisnejo-Gallo-2023-perturbative, Briozzo-Gallo-2023, Kumar-Beniamini-2023, BK-Tsupko-2023-time-delay, Sun-Er-Tsupko-2023, Feleppa-2024, Balek-2024, Comberiati-2025, Perlick-2025}, while earlier studies can be found in \cite{Synge-1960, Muhleman-1966, Muhleman-1970, Bicak-Hadrava-1975, Breuer-Ehlers-1980, Breuer-Ehlers-1981, Breuer-Ehlers-1981-AA, Kichenassamy-Krikorian-1985, Bliokh-Minakov-1989, Kulsrud-Loeb-1992, Krikorian-1999, Perlick-2000, Broderick-Blandford-2003, Broderick-Blandford-2003-ASS, Broderick-Blandford-2004}.
In particular, the boundary of a black hole's shadow is modified when the black hole is surrounded by plasma, since the plasma affects light rays through refraction. The resulting changes in the shadow's size and shape in the presence of cold plasma are discussed in detail in \cite{Perlick-Tsupko-BK-2015, Perlick-Tsupko-2017, BK-Tsupko-2017-Universe, Huang-2018, Yan-2019, Chowdhuri-2021-shadow-expand, Badia-Eiroa-2021, Perlick-Tsupko-2022, Bezdekova-2022, Li-2022, Badia-Eiroa-2023, Briozzo-Gallo-Madler-2023, Zhang-2023, Perlick-Tsupko-2024, Kobialko-2024, Vertogradov-2025, Kobialko-2025}. For recent calculation of photon rings in presence of plasma, see~\cite{Kobialko-2025, Frost-2025}.

Most studies dealing with this topic have focused on the model of a cold plasma, in which the refractive index takes a particularly simple form, allowing many results to be derived analytically. A more general case involving an arbitrary refractive index has been explored, e.g., in \cite{Tsupko-2021, Bezdekova-2023, Bezdekova-2024}. A notable property of cold plasma is that light ray propagation in this medium is independent of the medium's velocity \cite{Perlick-2000, Perlick-Tsupko-BK-2015, Schulze-Koops-Perlick-Schwarz-2017, Bezdekova-2024}. This property, however, does not hold for more general media. Therefore, by studying the effects of the medium's motion on the light-ray trajectories, we can probe deviations from the cold plasma model.

Motivated by this, in our previous study \cite{Bezdekova-2024} we analyzed the deflection of light by a spherically symmetric gravitating object surrounded by a moving, refractive, and dispersive medium -- going beyond the cold plasma case. We considered a refractive index of such a form that, first, allowed the medium’s velocity to influence light propagation (unlike in cold plasma) and, second, permitted a fully analytical treatment. Two specific cases of medium's motion were examined: radial motion and rotation in the equatorial plane, for which we calculated the deflection angles in the integral forms. Furthermore, we analyzed light deflection under the assumption that the refractive index constitutes a small perturbation from the cold plasma case.

In the present paper, we extend this analysis to axially symmetric spacetimes filled with a moving medium. Our motivation stems from the potential interplay between rotational effects originating from the spacetime and those induced by the motion of the medium.
The considered medium is refractive and dispersive, and within the geometrical optics approximation, we characterize it by its refractive index and velocity. The velocity of the medium is taken as a combination of radial motion and rotation. To carry out the analysis fully analytically and to obtain explicit integrable expressions, we adopt a specific -- yet sufficiently general -- functional form of the refractive index. This choice allows us to derive a closed-form expression for the deflection angle in the general asymptotically flat, axially symmetric case.

Subsequently, to enable a more detailed analytical investigation of the superposition of spacetime rotation and medium motion, we introduce an additional approximation: a slowly moving medium. This perturbative approach reveals two particularly interesting results. First, we demonstrate that -- to the first order in the medium velocity -- radial motion of the medium has no influence on the light propagation. Second, again to the first order, we show that for slowly rotating spacetimes, there exist conditions under which the rotational effects of the medium precisely cancel these of the central object. This quantifies how the rotational effects of a spacetime and the rotational effects of a medium interact. We emphasize that while our analysis employs several assumptions about the medium's motion and refractive index (enabling closed-form solutions), we impose no weak-field approximation or small-angle restriction on the gravitational light bending itself.

The paper is organized as follows. In Section \ref{sec:previous}, we review previous studies of the light propagation in axially symmetric spacetimes in the presence of a plasma. Unlike prior works discussed in Section \ref{sec:previous}, our analysis incorporates both the motion of the medium and a refractive index that goes beyond the cold plasma model. In Section \ref{sec:synge}, we begin by recalling the Synge’s formalism -- a standard framework for describing geometrical optics in the presence of both gravity and plasma. We then derive the equations of motion for the axially symmetric case with a general (unspecified) refractive index. Section \ref{sec:AxDefAng} introduces a specific (yet still broad) form of the refractive index. We derive the orbit equation for a medium undergoing both radial motion and rotation (Subsection \ref{ssec:mixedCase}). Next, we examine the special case of a purely radial motion, which yields a closed-form expression for the deflection angle (Subsection \ref{ssec:radialMotion}). The same analysis is then applied to a purely rotational case (Subsection \ref{ssec:rotatingMotion}). Then, in Section \ref{sec:slowmedium}, we consider the case of a slowly moving medium, where we find the surprising result that the radial medium motion does not contribute up to the first order in the orbit equation and show how rotational effects of the spacetime and of the medium cancel. In Section \ref{sec:Kerr}, we examine the important illustrative example of an axially symmetric spacetime: slowly rotating Kerr metric filled with a slowly moving medium. Our conclusions and discussion are given in Section \ref{sec:discussion}.

The metric signature applied in this paper reads $\{-,+,+,+\}$. The index convention is such that $i, k = 0,1,2,3$, resp. $(t,r,\vartheta, \varphi)$, and $\alpha, \beta = 1,2,3$, resp. $(r,\vartheta, \varphi)$. We set $G=c=1$.

\section{Previous studies of light propagation in axially symmetric spacetimes in presence of cold plasma} \label{sec:previous}

In this section, we present a short overview of studies of gravitational lensing in cold plasma and other types of media in an axially symmetric stationary spacetime, in particular in the Kerr metric. For a review of other studies of gravitational lensing in plasma, see, e.g., Ref.~\cite{BK-Tsupko-2017-Universe} and Sec.~2 of Ref.~\cite{BK-Tsupko-2023-time-delay}.

The first calculations of the light ray deflection near a Kerr black hole in the presence of plasma were presented in a monograph of Perlick \cite{Perlick-2000}. There, the motion of rays in the equatorial plane of the Kerr black hole surrounded by a spherically symmetric distribution of the cold plasma was considered, and the deflection angle in the form of an integral formula was derived. In the approximation of weak deflection and homogeneous plasma, gravitational lensing by a Kerr black hole was further investigated by Morozova et al. \cite{Morozova-2013}. For preceding studies of the Schwarzschild case in the weak deflection approximation, see Bisnovatyi-Kogan and Tsupko \cite{BK-Tsupko-2009, BK-Tsupko-2010}.

Crisnejo et al. \cite{Crisnejo-Gallo-Jusufi-2019} applied the Gauss-Bonnet theorem to calculate the deflection angles in stationary axially symmetric spacetimes in the presence of a cold plasma. In particular, they derived the deflection angle for light rays in the equatorial plane of the Kerr metric in the weak deflection case, including higher-order terms. To study the related works of this group with other interesting results, see also \cite{Crisnejo-Gallo-2018, Crisnejo-Gallo-Rogers-2019, Crisnejo-Gallo-Villanueva-2019, Briozzo-Gallo-2023, Briozzo-Gallo-Madler-2023}. Bezd\v{e}kov\'{a} and Bi\v{c}\'{a}k \cite{Bezdekova-2023} studied the deflection angle in the equatorial plane around an axially symmetric stationary object immersed in a general static dispersive medium characterized solely by its refractive index. Within this setting, they derived an integral expression for the deflection angle and applied it to the case of a cold plasma in the Hartle-Thorne metric. Additionally, they performed a weak deflection analysis and compared results between the Hartle-Thorne and Kerr spacetimes.

The propagation of light rays in plasma in the Kerr metric in arbitrary directions off the equatorial plane has been investigated in detail in a series of papers by Perlick and Tsupko \cite{Perlick-Tsupko-2017, Perlick-Tsupko-2024}. In \cite{Perlick-Tsupko-2017}, it was found that the separability of the Hamilton-Jacobi equation is possible only for special forms of the plasma distribution. Accordingly, for such distributions one can get equations of motion in the form of first-order differential equations. As a particular application, it becomes possible to calculate the shape of a black hole shadow analytically. In \cite{Perlick-Tsupko-2024}, the influence of a plasma on different types of orbits were analyzed: spherical, conical, circular. In particular, it was demonstrated that, unlike in vacuum, circular orbits can exist off the equatorial plane within the domain of outer communication of a Kerr black hole. Circular light rays in a cold plasma on an axially symmetric and stationary spacetime have been further studied in detail by Perlick \cite{Perlick-2025}.

The separability of the Hamilton-Jacobi equation and the calculation of a black hole shadow in cold plasma were further explored for a broader class of rotating black holes obtained by applying the Newman-Janis algorithm, see Bad\'ia and Eiroa \cite{Badia-Eiroa-2021}.  General axisymmetric and stationary spacetimes in the presence of a cold plasma medium were investigated analytically in Bezd\v{e}kov\'{a} \textit{et al.} \cite{Bezdekova-2022}. They derived the separability conditions on both the spacetime metric and plasma density, and obtained a general analytical expression for the resulting black hole shadow.

Having briefly summarized the state of the art of research on the light propagation in plasma in axially symmetric spacetimes, we will now proceed by introducing the mathematical formalism that serves as a foundation for our analysis of the light propagation in plasma.

\section{Synge's formalism and equations of motion with general refractive index}\label{sec:synge}

In this section, we review the Synge's formalism \cite{Synge-1960} and derive the equations that describe the motion of the light rays in the equatorial plane of an axially symmetric, stationary metric filled with a moving medium. At this stage, we do not specify the refractive index, aiming to proceed as far as possible at this purely general level instead. In the following Section~\ref{sec:AxDefAng}, we will consider a particular form of the refractive index, which will enable us to proceed much further and derive an explicit expression for the deflection angle. In Section~\ref{sec:slowmedium} we will apply the formalism presented here to derive the influence of slowly rotating media on light propagation.

To study light propagation within the geometrical optics approximation, we use the formalism introduced by Synge \cite{Synge-1960}. The Hamiltonian describing the motion of light rays in a transparent, isotropic medium within a gravitational field reads in canonical phase space coordinates $x^i$ and $p_i$
\begin{equation} \label{Hamiltonian-1}
\mathcal{H}(x,p) = \frac{1}{2} \left\{g^{ik} p_i p_k - \left[ n^2(x,\omega(x,p)) -1 \right] \left(p_j V^j \right)^2 \right\}  \, ,
\end{equation}
and we require
\begin{equation}
\mathcal{H}(x,p) = 0 \, .
\end{equation} 
The gravitational field is given by the spacetime metric $g^{ik}$, while the medium is characterized by its refractive index $n = n(x,\omega)$ and its 4-velocity $V^j=V^j(x)$ satisfying 
\begin{equation} \label{eq:norm}
g_{ij} V^i V^j = - 1 \, .
\end{equation}
The photon frequency $\omega$, measured in the rest frame of the medium, is given by
\begin{equation} \label{synge-freq-general}
\omega(x,p) = - \, p_i V^i(x)  \, .
\end{equation}
To perform calculations in a particular medium, one must specify its refractive index $n$ (as a scalar function of $x^i$ and $\omega$) and its 4-velocity $V^j$ (as a function of $x^i$) and substitute them into the Hamiltonian (\ref{Hamiltonian-1}). In a non-dispersive medium, the refractive index depends solely on spacetime coordinates $x^i$, while in a dispersive medium it is also a function of the photon frequency (\ref{synge-freq-general}). It should be emphasized that Synge's method, as presented in \cite{Synge-1960}, is not applicable to anisotropic media, for example, in the case of a medium with the presence of a magnetic field. We also note that in the chosen approach (geometrical optics), a ray is defined as the history of a particle -- photon. For this reason, these two concepts (rays and photons) can be used interchangeably. For more details, see \cite{Bicak-Hadrava-1975}.

The Hamiltonian formalism is favorable due to a straightforward derivation of the equations of motion. Indeed, the propagation of the light rays can be derived from the Hamilton's equations
\begin{equation} \label{Ham-equations}
\dot{x}^i = \frac{\partial \mathcal{H}}{\partial p_i}  \, , \; \; \dot{p}_i  = - \frac{\partial \mathcal{H}}{\partial x^i} \, .
\end{equation}
The overdot in $\dot{x}^i$ and $\dot{p}_i$ denotes the derivative with respect to the curve parameter.\vspace{11pt}

In this paper, we focus on axially symmetric stationary spacetimes, whose geometry is determined by a spacetime metric of the form
\begin{align}\label{metric_def4D}
        ds^2 &= g_{ik} dx^i dx^k = - A(r,\vartheta)dt^2 + B(r,\vartheta) dr^2\\
        & + 2P(r,\vartheta) \, dt \, d\varphi + D(r,\vartheta)d\vartheta^2 + C(r,\vartheta) d\varphi^2 \, . \nonumber
\end{align}
The components of inverse metric have the form:
\begin{align}\label{metric_def4Dinv}
  g^{rr} &= \frac{1}{B(r,\vartheta)},\quad g^{\vartheta \vartheta} = \frac{1}{D(r,\vartheta)}, \\
  g^{\varphi \varphi} &= \frac{A(r,\vartheta)}{A(r,\vartheta) \, C(r,\vartheta)+P^2(r,\vartheta)}, \nonumber\\
  g^{tt} &= \frac{-C(r,\vartheta)}{A(r,\vartheta) \, C(r,\vartheta)+P^2(r,\vartheta)}, \nonumber\\
  g^{t\varphi} &= \frac{P(r,\vartheta)}{A(r,\vartheta) \, C(r,\vartheta) + P^2(r,\vartheta)}.\nonumber
\end{align}
See also the discussion of this metric in the Appendix of~\cite{Bezdekova-2022}. The spherically symmetric case was considered in our previous paper \cite{Bezdekova-2024}, and can be reconstructed by setting $P(r,\vartheta)\to0$, 
$A(r,\vartheta) \to A(r)$, $B(r,\vartheta) \to B(r)$, and $C(r,\vartheta)\to D(r) \sin^2\vartheta$, $D(r,\vartheta)\to D(r)$.

In order to derive the orbit equation in the equatorial plane consistently, we need to make additional assumptions about the metric coefficients $A$, $B$, $C$, $D$, $P$, the refractive index of the medium $n$, and the velocity profile of the medium $V=(V^t,V^r,V^\vartheta,V^\varphi)$ -- that is, about all functions from which the Hamiltonian \eqref{Hamiltonian-1} is constructed.
To ensure the existence of an equatorial plane $(\vartheta = \pi/2, p_\vartheta = 0)$ and that light rays remain confined to it, we assume that the dependence of the total Hamiltonian on $\vartheta$ is symmetric around $\vartheta = \pi/2$, i.e., $\mathcal{H}|_{\vartheta = \pi/2+a}=\mathcal{H}|_{\vartheta = \pi/2-a}$ for any $a\in(0,\pi/2)$; see also \cite{Bezdekova-2022, Perlick-Tsupko-2024}. Furthermore, to guarantee that $p_0$ and $p_\varphi$ are constants of motion, we assume that the total Hamiltonian \eqref{Hamiltonian-1} does not depend on $t$ or $\varphi$. Finally, we restrict to a medium whose motion is radial and rotational by setting $V^\vartheta=0$.

Under these assumptions, we can restrict \eqref{Hamiltonian-1} with \eqref{metric_def4D} to the equatorial plane $(\mbox{i.e., to set }\vartheta = \pi/2, p_\vartheta = 0)$ and suppress the $\vartheta$-dependence of the functions $A,B,C,D,P$, the refractive index $n$ and the velocity 
\begin{align}\label{velocity_mixed}
    V=(V^t,V^r,0,V^\varphi)
\end{align} 
in the following. The metric in the equatorial plane is given by 
\begin{align}\label{metric_def}
    ds^2|_{\vartheta=\frac{\pi}{2}} &=  - A(r)dt^2 + B(r) dr^2\\
        & + 2P(r) \, dt \, d\varphi  + C(r) d\varphi^2 \,, \nonumber
\end{align}
while the photon frequency becomes 
\begin{equation}\label{omega_mixed}
\omega = \omega(p_r,p_\varphi,r) = - p_0 V^0(r) - p_r V^r(r) -p_\varphi V^\varphi(r)  \, .
\end{equation}
We obtain the Hamiltonian \begin{align} 
\label{eq:generalHamiltonian}
    \mathcal{H} = \frac{1}{2} &\Big\{\frac{p_r^2}{B(r)} + \frac{A(r) p_\varphi^2 - p_0^2 C(r) + 2 p_0 p_\varphi P(r)}{A(r) C(r)+P^2(r)} \\
   &+ w(\omega,r) \Big\} \,, \nonumber
\end{align}
where the last term $w$ is introduced in the same way as in \cite{Bezdekova-2024} and reads
\begin{equation}
w(\omega,r) = - \, (n^2(\omega,r) - 1) \, \omega^2 \, ,
\end{equation}
with $\omega$ in the form \eqref{omega_mixed}.

The term $w(\omega,r)$ is convenient to use in situations where the refractive index is not explicitly specified, as it allows all formulas to be written in a more compact form. In particular cases, the function $w$ simplifies accordingly: in vacuum, $n=1$, it is zero, $w=0$, and in the case of a cold plasma,
\begin{equation} \label{refr-index-plasma-r}
n^2(r, \omega) = 1 - \frac{\omega_p^2(r)}{\omega^2} \, ,
\end{equation}
it equals the squared plasma frequency, $w = \omega_p^2(r)$, see \cite{Bezdekova-2024} for details.

Since the Hamiltonian \eqref{eq:generalHamiltonian} is independent of both time $t$ and the azimuthal angle $\varphi$, the equations of motion
\begin{equation}\label{eq:cons}
\dot{p}_0 = 0 \, , \quad \dot{p}_\varphi = 0 \, ,
\end{equation}
imply that the quantities $p_0$ and $p_\varphi$ are constants of motion (in addition to $\mathcal{H}$). Moreover, by additionally restricting our analysis to an asymptotically flat metric, we find that the constant 
$p_0$ is equal (up to a sign) to the photon frequency at infinity:
\begin{equation}\label{eq:p0w0}
p_0 = - \omega_0 \, . 
\end{equation}
This will be used further, e.g., when applying in Eq.~\eqref{omega_mixed}. We emphasize that without this additional assumption -- i.e., if the metric were not asymptotically flat -- $p_0$ would still remain a constant of motion. However, it would no longer correspond to the photon frequency at infinity and thus would lack a straightforward physical interpretation.

The equations of motion in the equatorial plane of the metric (\ref{metric_def}), for a moving medium with refractive index of the form $n=n(\omega,r)$ and 4-velocity \eqref{velocity_mixed}, follow from the Hamiltonian \eqref{eq:generalHamiltonian} as:
\begin{align} \label{eq:dot-r-general}
        \dot{r} 
        &= \frac{\partial\mathcal{H}}{\partial p_r}\\
        &=\frac{p_r}{B(r)}+ \frac{1}{2} \frac{\partial w}{\partial \omega} \frac{\partial \omega(p_r,p_\varphi,r)}{\partial p_\varphi}\nonumber\\
        &= \frac{p_r}{B(r)} -\frac{1}{2} \frac{\partial w}{\partial \omega} V^r(r) \, , \nonumber
\end{align}
and
\begin{align} \label{eq:dot-phi-general}
        \dot{\varphi}&= \frac{\partial\mathcal{H}}{\partial p_\varphi}\\
        &=\frac{p_\varphi A(r)-\omega_0 P(r)}{A(r) C(r)+P^2(r)}+ \frac{1}{2} \frac{\partial w}{\partial \omega} \frac{\partial \omega(p_r,p_\varphi,r)}{\partial p_\varphi}\nonumber\\ 
        &=\frac{p_\varphi A(r)-\omega_0 P(r)}{A(r) C(r)+P^2(r)} - \frac{1}{2} \frac{\partial w}{\partial \omega}  V^\varphi(r) \, . \nonumber
\end{align}
Note that in both equations of motion there appear terms which stem from the dependence of the wave frequency~\eqref{omega_mixed} on $p_r$ and $p_\varphi$  (cf. Section~\ref{sec:slowmedium}). We can write the orbit equation in the form:
\begin{align} \label{eq:orbit-eq-prelim}
        \frac{d\varphi}{dr}=
        &\left(\frac{p_\varphi A(r)-\omega_0 P(r)}{A(r) C(r)+P^2(r)}- \frac{1}{2} \frac{\partial w}{\partial \omega} V^\varphi(r)\right)\\
        &\times\left(\frac{p_r}{B(r)} -\frac{1}{2} \frac{\partial w}{\partial \omega} V^r(r)\right)^{-1}.\nonumber
\end{align}
It is important to note that, at this general level, the components $V^r$ and $V^\varphi$ enter the orbit equation (\ref{eq:orbit-eq-prelim}) in different ways. Specifically, the dependence on $V^\varphi$ occurs only in the numerator, whereas the dependence on $V^r$ enters the expression solely in the denominator.
The difference comes from the definition \eqref{omega_mixed} and the form of the equations of motion. This feature has further consequences which are discussed in Section~\ref{sec:slowmedium}.

Eq.~\eqref{eq:orbit-eq-prelim} does not represent the final form of the orbit equation, but rather serves as an intermediate step in its derivation. While $p_\varphi$ and $\omega_0$ are constants of motion that characterize a particular orbit, the quantity $p_r$ remains unknown at this stage. To proceed with the calculations, it is necessary to express $p_r$ in terms of the other variables and substitute it into the equations of motion \eqref{eq:dot-r-general}, \eqref{eq:dot-phi-general}, or the orbit equation \eqref{eq:orbit-eq-prelim}. This can be achieved, for example, by solving $\mathcal{H}=0$ for $p_r$.

However, an explicit form of $p_r$ cannot be obtained without further specifying the refractive index, since, in general, $p_r$ appears within the frequency \eqref{omega_mixed} and, consequently, within the function $w$. Moreover, such an expression can be derived analytically only for certain specific forms of $n$. Therefore, in the next section, we choose a particular form of the refractive index in order to determine the deflection angle in selected cases. Subsequently, we will study this quantity for a slowly moving medium and slowly rotating Kerr spacetime in Sections \ref{sec:slowmedium} and \ref{sec:Kerr}, respectively.

\section{Deflection Angle of Light Ray for a Particular Choice of Refractive Index} \label{sec:AxDefAng}

As explained in the previous section, deriving an explicit formula for the deflection angle requires either choosing a specific form of the refractive index, rather than working with the general expression $n=n(\omega, r)$, or resorting to perturbative regimes (as we will do in Section \ref{sec:slowmedium}). In this section, we use the following form of the refractive index: 
\begin{equation}\label{refr_n}
n^2(\omega, r) = a_0(r) + \frac{a_1(r)}{\omega} + \frac{a_2(r)}{\omega^2} \, ,
\end{equation}
where $a_0(r)$, $a_1(r)$, and $a_2(r)$ are functions of $r$ only, and $\omega = \omega(p_r, p_\varphi, r)$ is given by Eq.~(\ref{omega_mixed}). 
The coefficients can be chosen with considerable freedom; the only requirements are that the refractive index remains positive and that the corresponding group velocity of light in the medium remains less than the speed of light in vacuum. In physical situations they are determined by the properties of the medium under consideration.

This particular form was previously used by us in Ref.~\cite{Bezdekova-2024}. From a mathematical point of view, the advantage of using the refractive index (\ref{refr_n}) lies in the fact that the condition $\mathcal{H} = 0$ becomes a quadratic equation in the variable $p_r$, making it analytically tractable. From a physical perspective, this refractive index includes two important particular cases: when $a_1=a_2=0$, it corresponds to a non-dispersive medium; and when $a_0=1$, $a_1=0$ with $a_2 = - \omega_p^2(r)$ it describes a cold plasma.

In this section, the choice of a specific refractive index allows us to derive the final form of the orbit equation. This equation can serve as the integrand for the deflection angle and, consequently, can be used to compute the deflection angle itself (see below).

In Subsection~\ref{ssec:mixedCase}, we consider a medium that experiences both radial motion and rotation in the equatorial plane. However, the resulting expression for the deflection angle integrand is rather complex and poses difficulties for the analysis, especially for the comparison with the results of our previous work~\cite{Bezdekova-2024}. For this reason, in the following subsections we consider the cases of pure radial motion (Subsection~\ref{ssec:radialMotion}) and pure rotational motion (Subsection~\ref{ssec:rotatingMotion}) separately. 
Due to the analytical challenges presented by the general case of refractive index, we further investigate it perturbatively, for a slowly moving medium, in Section~\ref{sec:slowmedium}.

\begin{widetext}
    
\subsection{Orbit Equation for a Medium with Both Radial and Rotational Motion} \label{ssec:mixedCase}

In the case of simultaneous radial motion and rotation in the equatorial plane, the 4-velocity of the medium and the wave frequency are given by Eqs.~(\ref{velocity_mixed}) and (\ref{omega_mixed}), respectively. The refractive index is given by \eqref{refr_n}.

Upon substituting \eqref{refr_n} into the Hamiltonian \eqref{eq:generalHamiltonian}, we find:
\begin{gather}\label{Hamiltonian_Sec4a}
  \mathcal{H}
  =\frac{1}{2} \, \bigg\{ \frac{p_r^2}{B(r)}
  +\frac{A(r) p_\varphi^2-\omega _0^2 C(r)-2 \omega_0 p_\varphi P(r)}{A(r) C(r)+P^2(r)} - a_2(r)+
  \left( V^0 \omega _0-V^r(r) p_r-V^\varphi(r)p_\varphi\right) \\
  \times \left[ (1-a_0(r))(V^0 \omega _0-V^r(r) p_r-V^\varphi(r)p_\varphi)-a_1(r)\right]
  \bigg\} \nonumber.
\end{gather}
The $V^0$ component of the medium's 4-velocity can be expressed in terms of $V^r$ and $V^\varphi$ components using the normalization condition \eqref{eq:norm} as follows
\begin{equation} \label{eq:V0gen}
V^0(r) = \frac{1}{A(r)} \left\{V^\varphi(r) P(r) + \sqrt{(V^\varphi(r))^2[A(r)C(r)+P^2(r)]+A(r)[B(r)(V^r(r))^2+1]}\right\}.
\end{equation}
Formally, a plus-minus sign should appear in front of the square root, indicating the existence of two solutions -- one positive and one negative. However, in accordance with previous studies \cite[e.g.,][]{Bezdekova-2024}, only the solution with the plus sign is considered. Since $V^0(r)$ is the derivative of the time coordinate with respect to the curve parameter, positive $V^0(r)$ (solution with plus sign) corresponds to future-oriented motion, where time and curve parameter have the same sign. This choice is preferably applied without the loss of generality.

The expression \eqref{eq:V0gen} for $V^0(r)$ demonstrates one of the main differences between the spherically symmetric case studied in \cite{Bezdekova-2024} and the axially symmetric metric considered in the present paper. The presence of the non-vanishing off-diagonal component $P(r)$ in the metric makes the expression \eqref{eq:V0gen} severely more involved. Although we have presented the explicit expression for $V^0(r)$, we will retain the symbolic form $V^0(r)$ in the following formulas. This is done solely for aesthetic reasons, to avoid unnecessarily lengthy expressions. Throughout, $V^0(r)$ should always be understood as given in \eqref{eq:V0gen}.

The Hamiltonian (\ref{Hamiltonian_Sec4a}) is quadratic in both $p_r$ and $p_\varphi$. For practical calculations, the dependence on $p_r$ is more under scope. This is because $p_r$ is not a constant of motion and its dependence on the remaining variables must be expressed explicitly by solving the condition $\mathcal{H}=0$. The described behavior of $p_r$ allows one to rewrite the Hamiltonian (\ref{Hamiltonian_Sec4a}) in the form convenient for further calculations as
\begin{equation} \label{eq:H-quadratic}
\mathcal{H}= \frac{1}{2} \left[ \mathcal{A}_r(r)p_r^2 + 2\mathcal{B}_r(r)p_r + \mathcal{C}_r(r,p_\varphi) \right],
\end{equation}
where
\begin{align}
    &\mathcal{A}_r(r) = \frac{1}{B(r)} + (1-a_0(r)) (V^r(r))^2 \, , \label{eq:A_r-general} \\
    &\mathcal{B}_r(r) = V^r(r) \label{eq:B_r-general} \Big[ (a_0(r)-1) \, (\omega_0 V^0(r)-p_\varphi V^\varphi(r)) + \frac{1}{2}a_1(r) \Big]  \, , \\
    &\mathcal{C}_r(r,p_\varphi) = \frac{A(r) p_\varphi^2-\omega _0^2 C(r)-2 \omega_0 p_\varphi P(r)}{A(r) C(r)+P^2(r)}+\mathcal{C}_{r1}(r,p_\varphi) \, , \label{C_r-general} \\
&\mathcal{C}_{r1}(r,p_\varphi) =  - \, a_2(r)+ \, (\omega_0 V^0(r)-p_\varphi V^\varphi(r)) \label{Cr1_gen} \left[(1-a_0(r)) \, (\omega_0 V^0(r)-p_\varphi V^\varphi(r))-a_1(r)\right] \, .
\end{align}
At this stage, it is instructive to compare the coefficients \eqref{eq:A_r-general}, \eqref{eq:B_r-general}, \eqref{C_r-general}, and \eqref{Cr1_gen} with those in Section IV of our previous paper \cite{Bezdekova-2024}, where a spherically symmetric spacetime with a radially moving medium was analyzed. The present discussion addresses a more general situation: first, the background metric is rotating (i.e.,\ axially symmetric) rather than spherically symmetric; and second, both radial and azimuthal components of the medium’s motion are considerred at the same time. The function $\mathcal{A}_r(r)$ is the same as was introduced in \cite{Bezdekova-2024}. The functions $\mathcal{B}_r(r)$ and $\mathcal{C}_r(r,p_\varphi)$ are now more complicated as the wave frequency \eqref{omega_mixed} includes more terms, and the term $\mathcal{C}_{r1}(r,p_\varphi)$ is additionally a function of momentum $p_\varphi$, which was not the case in \cite{Bezdekova-2024}. 

The equations of motion \eqref{eq:dot-r-general} and \eqref{eq:dot-phi-general} in this case return 
\begin{align}
\dot{r} &=  \mathcal{A}_r(r) \, p_r + \mathcal{B}_r(r) \, ,\\
\label{eq:phi-motion-case-a}
\dot{\varphi}&=\frac{A(r) p_\varphi-\omega_0 P(r)}{A(r) C(r)+P^2(r)}+V^\varphi(r) \Big\{ (a_0(r)-1)  \left[ \omega_0 V^0(r)-p_\varphi V^\varphi(r)-V^r(r) \right] +\frac{1}{2}a_1(r) \Big\}.
\end{align}
The momentum component $p_r$ can be found from $\mathcal{H}=0$ as
\begin{equation}
    p_r = \frac{-\mathcal{B}_r(r) \pm\sqrt{\mathcal{B}_r^2(r)-\mathcal{A}_r(r)\mathcal{C}_r(r,p_\varphi)}}{\mathcal{A}_r(r)} \, .
\end{equation}

With all this at hand, one can express the orbit equation as
\begin{align}
\frac{d\varphi}{dr} =
&\pm \left\{\frac{A(r) p_\varphi-\omega_0 P(r)}{A(r) C(r)+P^2(r)} + V^\varphi(r) \Big[(a_0(r)-1) \, (\omega_0 V^0(r)-p_\varphi V^\varphi(r)-V^r(r))+\frac{1}{2}a_1(r)\Big] \right\} \\
&\times\left[\mathcal{B}_r^2(r)-\mathcal{A}_r(r) \mathcal{C}_r(r,p_\varphi) \right]^{-1/2} \,.\nonumber
\end{align}
By explicitly substituting $\mathcal{A}_r(r)$, $\mathcal{B}_r(r)$, $\mathcal{C}_r(r)$ from Eqs.~\eqref{eq:A_r-general}, \eqref{eq:B_r-general}, \eqref{C_r-general}, and performing some algebraic manipulations, we obtain the final form of the orbit equation:
\begin{align}\label{integ_mixed_n}
\frac{d\varphi}{dr} =& \pm\frac{\sqrt{A(r)B(r)}}{\sqrt{A(r) C(r)+P^2(r)}}\\
&\times \left\{ \frac{p_\varphi}{\omega_0}-\frac{P(r)}{A(r)} +\frac{A(r) C(r)+P^2(r)}{A(r)\omega_0}V^\varphi(r) \Big[ (a_0(r)-1) \, (\omega_0 V^0(r)-p_\varphi V^\varphi(r)-V^r(r))+\frac{1}{2}a_1(r) \Big] \right\} \nonumber\\
&\times \bigg\{ \bigg[ \frac{C(r)}{A(r)}+\frac{P^2(r)}{A^2(r)} - \left( \frac{p_\varphi}{\omega _0}
-\frac{P(r)}{A(r)} \right)^2 \, \bigg] \left(1+(1-a_0(r))B(r)(V^r(r))^2\right) \nonumber\\
&+ \frac{A(r) C(r)+P^2(r)}{\omega^2_0A(r)}
\left[B(r)(V^r(r))^2 \Big( a_2(r)(1-a_0(r))+ \frac{1}{4}a^2_1(r) \Big) - \mathcal{C}_{r1}(r,p_\varphi)\right]\bigg\} ^{-1/2}.\nonumber
\end{align}
\end{widetext}

The right-hand side of the orbit equation \eqref{integ_mixed_n} can serve as the integrand for the deflection angle. To obtain the deflection angle, this expression must be integrated with respect to $r$ over the appropriate limits; see the next subsections, \ref{ssec:radialMotion} and \ref{ssec:rotatingMotion}, for details.

The right-hand side of Eq.~\eqref{integ_mixed_n} depends on multiple variables. The functions $A(r)$, $B(r)$, $C(r)$, $P(r)$ define the metric \eqref{metric_def}; the coefficients $a_0(r)$, $a_1(r)$, $a_2(r)$ characterize the refractive index as given in Eq.~\eqref{refr_n}; and the functions $V^0(r)$, $V^r(r)$, $V^\varphi(r)$ describe the motion of the medium. Additionally, there are two external parameters that define a particular light ray: the constants of motion $\omega_0$ and $p_\varphi$. The plus-minus sign in front of the entire right-hand side must be selected according to the signs of $d\varphi$ and $dr$ along each segment of the ray trajectory.

In contrast to the vacuum case, where a particular light ray is characterized by a single external parameter, light propagation in a medium requires two. For the purpose of deriving the deflection angle, it is more convenient to switch from $p_\varphi$ to the distance of closest approach $R$, defined as the minimum value of the radial coordinate along the trajectory. This choice is natural, as the deflection angle integral already includes $R$ as one of its integration limits; see, e.g., Subsection \ref{ssec:radialMotion}. On basis of that, the typical next step would be to eliminate the ratio  $p_\varphi/\omega_0$ by solving the condition $d\varphi/dr=0$ at $r=R$. However, in the case of Eq.~\eqref{integ_mixed_n}, this would require to solve a quadratic equation for $p_\varphi/\omega_0$ with many terms and for which no obvious convenient combinations which could lead to simplifications could be found. Nevertheless, Eq.~\eqref{integ_mixed_n} remains valuable for analyzing the influence of individual velocity components and serves as a starting point for the simplified cases of medium motion discussed in Subsections \ref{ssec:radialMotion} and \ref{ssec:rotatingMotion}. In particular, it is not necessary to explicitly solve for $p_r$ again in the following subsections.

\subsection{Deflection Angle for a Radially Moving Medium} \label{ssec:radialMotion}
In this subsection, we consider a purely radially moving medium. This is a particular case of the previous subsection, occurring when 
$V^{\varphi}(r)=0$. Correspondingly, the medium’s velocity is given by:
\begin{equation}
V^i = (V^0(r),V^r(r), 0, 0) \, .
\end{equation}
Consideration of this particular case yields simpler expressions, enabling us to proceed further in comparison with Subsection~\ref{ssec:mixedCase} and obtain a closed-form expression for the deflection angle.

Note that in Sections III and IV of Ref.~\cite{Bezdekova-2024}, the radial velocity component $V^r(r)$ was renamed to $f(r)$. In this paper, we keep the notation $V^r(r)$ throughout in order to highlight the role of this specific velocity component more transparently. Apart from this change, we follow the same logic as discussed in \cite{Bezdekova-2024} and extend it to the case of axially symmetric spacetimes. The case $V^r(r)<0$ corresponds to spherically symmetric accretion, while $V^r(r)>0$ corresponds to a spherical outflow, such as a stellar wind around a compact object.

The $V^0$ component of the medium 4-velocity, in a general form given by (\ref{eq:V0gen}), reduces in this particular case ($V^\varphi=0$) to
\begin{equation}
V^0(r) = \sqrt{\frac{B(r) (V^r(r))^2+1}{A(r)}} \, .
\end{equation}
This formula is not used explicitly in what follows, but it is assumed that it can be substituted into subsequent expressions where needed. Moreover, the photon frequency \eqref{omega_mixed} in this case returns
\begin{equation}
\omega(p_r,r) = V^0(r) \, \omega _0 - V^r(r) \, p_r \, .
\end{equation}

In the orbit equation (\ref{integ_mixed_n}), one can now set $V^\varphi(r)=0$ to get
\begin{widetext}
    \begin{align}
\frac{d\varphi}{dr} =& \pm\frac{\sqrt{A(r)B(r)}}{\sqrt{A(r) C(r)+P^2(r)}}
 \biggl( \frac{p_\varphi}{\omega_0} - \frac{P(r)}{A(r)} \biggr) \\
&\times \Biggl\{ \biggl[ \frac{C(r)}{A(r)}+\frac{P^2(r)}{A^2(r)} - \biggl( \frac{p_\varphi}{\omega _0}
-\frac{P(r)}{A(r)} \biggr)^{\!2} \; \biggr] \left(1+(1-a_0(r))B(r)(V^r(r))^2\right) \nonumber\\
& + \, \frac{A(r) C(r)+P^2(r)}{\omega^2_0A(r)}
\Big[ B(r)(V^r(r))^2 \Big( a_2(r)(1-a_0(r))+ \frac{1}{4}a^2_1(r) \Big) -\mathcal{C}_{r1}(r) \Big] \Biggr\}^{-1/2},\nonumber
\end{align}
where the expression \eqref{Cr1_gen} simplifies to
\begin{equation}
\mathcal{C}_{r1}(r) =  - \, a_2(r) + \omega_0 V^0(r)\left[(1-a_0(r)) \, \omega_0 V^0(r)-a_1(r)\right] \, .
\end{equation}

For further convenience, let us introduce the function $\tilde{\mathcal{B}}_r(r)$ (not to be confused with $\mathcal{B}_r(r)$ defined in Eq.~\eqref{eq:B_r-general}):
\begin{equation}
\tilde{\mathcal{B}}_r(r) = \frac{B(r)(V^r(r))^2\left(a_2(r)(1-a_0(r))+ \frac{1}{4}a^2_1(r) \right)-\mathcal{C}_{r1}(r)}{1 + B(r) (V^r(r))^2(1-a_0(r))}.
\end{equation}
This allows us to write
\begin{align}
\frac{d\varphi}{dr} = \pm \frac{\sqrt{A(r)}}{\sqrt{\mathcal{A}_r(r)(A(r) C(r)+P^2(r))}} \left\{ \frac{\frac{C(r)}{A(r)}+\frac{P^2(r)}{A^2(r)}+\frac{A(r) C(r)+P^2(r)}{\omega^2_0A(r)} \tilde{\mathcal{B}}_r(r)}{ \left(\frac{p_\varphi}{\omega_0}-\frac{P(r)}{A(r)} \right)^2}-1 \right\}^{-1/2} \, ,
\end{align}
where $\mathcal{A}_r(r)$ is defined in \eqref{eq:A_r-general}.
\end{widetext}

For further simplification, it is convenient to introduce the function $h(r)$ as
\begin{equation}\label{eq:h-radial-case}
  h^2(r)=\frac{A(r)C(r)+P^2(r)}{A^2(r)} \left[ 1 +\frac{A(r)}{\omega^2_0} \tilde{\mathcal{B}}_r(r) \right] \, .
\end{equation}
Similar functions, typically playing the role of an effective potential, have already been widely used -- first introduced by Perlick \cite{Perlick-2000} and later employed in related studies \cite[e.g.,][]{Perlick-Tsupko-BK-2015, Tsupko-2021, Bezdekova-2023, Perlick-Tsupko-2022, Bezdekova-2024}.
The function \eqref{eq:h-radial-case}, together with the condition $d\varphi/dr=0$ at the point of the closest approach $r=R$, allows us to express the ratio $p_\varphi/\omega_0$ as
\begin{equation} \label{eq:p-phi-omega0-hR}
\frac{p_\varphi}{\omega_0}=\frac{P(R)}{A(R)}\pm h(R) \, .
\end{equation}
This equation appears the same as Eq.~(19) in \cite{Bezdekova-2023}, but the definition of $h(r)$ differs and is more general here. Applying the relation between $p_\varphi/\omega_0$ and $h(r)$ yields
\begin{align} \label{eq:orbit-eq-radial-case}
\frac{d\varphi}{dr} =&\pm\frac{\sqrt{A(r)}}{\sqrt{\mathcal{A}_r(r)(A(r) C(r)+P^2(r))}}\\
&\times \Bigg\{ \frac{h^2(r)}{\big( \frac{P(R)}{A(R)}-\frac{P(r)}{A(r)}\pm h(R) \big)^2} - 1 \Bigg\}^{-1/2}.\nonumber
\end{align}

\begin{figure}
\begin{center}
\includegraphics[width=0.99\columnwidth]{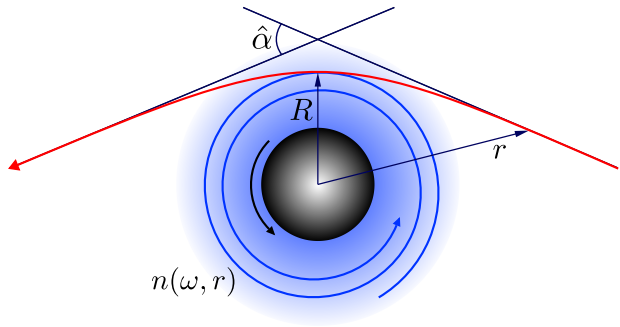}
\end{center}
\caption{Deflection angle $\hat{\alpha}$ of a light ray propagating in the equatorial plane of an axially symmetric spacetime filled with a moving medium of refractive index $n(\omega, r)$. The medium moves around the central object with both radial and azimuthal components of velocity. The light ray has radial coordinate $r$, travels from infinity, passes through the point of closest approach $r=R$ and continues back to infinity.}
\label{fig:deflection}
\end{figure}

We now proceed to calculate the deflection angle, a useful quantity for characterizing the light propagation, particularly relevant in gravitational lensing problems. We consider light rays in the equatorial plane of an asymptotically flat axially symmetric spacetime, which originate from infinity, approach a gravitating object, reach a minimum radial distance $R$, and then return to infinity (cf. Fig.~\ref{fig:deflection}). The deflection angle is defined as the angle between the asymptotic directions of the incoming and outgoing segments of the trajectory (see, e.g., \cite{Bozza-2010-review}). This definition is primarily applicable in asymptotically flat spacetimes, where such asymptotic behavior is well-defined.

The orbit equation \eqref{eq:orbit-eq-radial-case} contains a plus-minus sign in two places: one in front of the entire right-hand side, and another near the term $h(R)$. These signs must be chosen carefully during the calculation of the deflection angle; see Section V of Ref.~\cite{Bezdekova-2024} for comparison. For further convenience, let us rewrite Eq.~\eqref{eq:orbit-eq-radial-case} as
\begin{equation} \label{eq:radial-orbit-compact}
\frac{d\varphi}{dr} = \pm f_\pm(r) \, ,
\end{equation}
where the plus-minus in the subscript of $f_\pm(r)$ refers to the choice of sign in front of the term $h(R)$.

Let us first consider the case where the azimuthal coordinate 
$\varphi$ increases along the photon motion, i.e., $d\varphi>0$. Since $\dot{\varphi}>0$, from \eqref{eq:phi-motion-case-a} with $V^\varphi(r)=0$, we obtain
\begin{equation} \label{eq:ratio-plus}
\frac{A(r) p_\varphi-\omega_0 P(r)}{A(r) C(r)+P^2(r)} > 0 \, .
\end{equation}
The relation \eqref{eq:ratio-plus} holds for arbitrary $r$ of photon trajectory and thus also for $R$. Using this result in Eq.~\eqref{eq:p-phi-omega0-hR}, we find that the sign ``plus'' should be chosen there. Accordingly, the positive sign should also be selected in Eq.~\eqref{eq:orbit-eq-radial-case} near the term $h(R)$, and we will have $f_+$ in Eq.~\eqref{eq:radial-orbit-compact}.

We now consider the case where $d\varphi>0$; however, the sign of 
$dr$ changes during the photon’s trajectory. During the approach phase, when the photon moves towards the central object, we have 
$dr<0$, and therefore the negative sign should be selected in front of $f_+$ in Eq.~\eqref{eq:radial-orbit-compact}. Conversely, during the departure phase, when the photon moves away after reaching the closest distance $R$, $dr>0$, and the positive sign must be chosen.

The total change in the azimuthal coordinate $\varphi$ along the photon trajectory is given by:
\begin{align}
\Delta \varphi =& \int \limits_\infty^R  \frac{d\varphi}{dr} dr + \int \limits_R^\infty  \frac{d\varphi}{dr} dr \\
=& - \int \limits_\infty^R f_+(r) \, dr + \int \limits_R^\infty f_+(r) \, dr \nonumber \\
=& \; 2 \int \limits_R^\infty f_+(r) \, dr \, .\nonumber 
\end{align}

Now let us consider the case in which the azimuthal coordinate 
$\varphi$ decreases along the photon trajectory, i.e., $d\varphi<0$. 
An analogous line of reasoning leads to the conclusion that the minus sign should be selected in Eqs.~\eqref{eq:p-phi-omega0-hR} and \eqref{eq:orbit-eq-radial-case} in front of $h(R)$, and we will have $f_-$ in \eqref{eq:radial-orbit-compact}. With the appropriate choice of signs in front of $f_-(r)$ in the Eq.~\eqref{eq:radial-orbit-compact} during incoming and outcoming segments of the trajectory, we find:
\begin{equation}
\Delta \varphi = - 2 \int \limits_R^\infty f_-(r) \, dr \, .
\end{equation}

Defining the deflection angle $\hat{\alpha}$ as a positive quantity, such that $\pi$ (corresponding to a straight-line trajectory) is subtracted, we write:
\begin{equation}
    \hat{\alpha} = \pm \, \Delta \varphi - \pi \, .
\end{equation}
Here, the plus sign must be selected for orbits with $\dot{\varphi}>0$, while the minus sign applies for orbits with $\dot{\varphi}<0$. Finally, the deflection angle takes the form:
\begin{align} \label{eq:defl-angle-final}
    \hat{\alpha}  = \; & 2 \int \limits_R^\infty \frac{\sqrt{A(r)}}{\sqrt{\mathcal{A}_r(r)(A(r) C(r)+P^2(r))}}\\
&\times \Bigg\{ \frac{h^2(r)}{\big( \frac{P(R)}{A(R)} - \frac{P(r)}{A(r)} \pm h(R) \big)^2} - 1 \Bigg\}^{-1/2} dr \, - \, \pi \, . \nonumber
\end{align}
The deflection angle calculated in this manner is always positive, regardless of the direction of  $\varphi$-motion. One should use the ``plus'' sign in front of $h(R)$ in \eqref{eq:defl-angle-final} for motion with increasing $\varphi$ (i.e., $\dot{\varphi}>0$), and the minus sign when $\varphi$ is decreasing ($\dot{\varphi}<0$). For quick reference, we recall that the coefficients $A(r)$, $B(r)$, $C(r)$ and $P(r)$ are the metric functions (cf.~\eqref{metric_def}); the expression $\mathcal{A}_r(r)$ is given by \eqref{eq:A_r-general}; and the function $h(r)$ is defined by \eqref{eq:h-radial-case}. The deflection angle depends on two additional parameters: the closest approach distance $R$ and the photon frequency at infinity $\omega_0$; the latter being embedded in function $h(r)$. 

The deflection angle given by Eq.~\eqref{eq:defl-angle-final} describes light deflection in the equatorial plane of an axially symmetric stationary spacetime with a radially moving medium. By setting 
$V^r(r)=0$ in Eq.~\eqref{eq:defl-angle-final}, we recover the particular case of an axially symmetric stationary spacetime with a static medium, as derived in \cite{Bezdekova-2023}, see Eq.~(21) therein. Alternatively, by setting $P(r)=0$ in Eq.~\eqref{eq:defl-angle-final}, this formula reduces to the particular case of a spherically symmetric spacetime with a radially moving medium, presented in \cite{Bezdekova-2024}, see Eq.~(55) therein.

\subsection{Deflection Angle for Rotating Medium}\label{ssec:rotatingMotion}

In this subsection, we consider a purely rotating medium in the equatorial plane of axially symmetric stationary metric.
This is another particular case of Subsection~\ref{ssec:mixedCase}, which occurs when $V^r(r)=0$. Consequently, the velocity of the medium is given by
\begin{equation}
V^i = (V^0(r), 0, 0, V^\varphi(r))\,.
\end{equation}
Analogously to Subsection~\ref{ssec:radialMotion}, we do not rename 
$V^\varphi(r)$ to $f(r)$, as was done in Section~V of Ref.~\cite{Bezdekova-2024}, and instead retain $V^\varphi(r)$ in all subsequent expressions.

With the substitution $V^r(r)=0$, the normalization condition of the 4-velocity (\ref{eq:V0gen}) reduces to
\begin{equation}
V^0(r) = \frac{V^\varphi(r) P(r) + \sqrt{(A(r) C(r)+ P^2(r))(V^\varphi(r))^2+A(r)}}{A(r)} \, ,
\end{equation}
the photon frequency \eqref{omega_mixed} in this case reads
\begin{equation}\label{omega}
 \omega(p_\varphi,r)= V^0(r) \omega _0-V^\varphi(r) p_\varphi \, ,
\end{equation}
and Eq.~(\ref{integ_mixed_n}) becomes
\begin{widetext}
\begin{align}\label{integrand_rotating_n}
\frac{d\varphi}{dr} =& \pm\frac{\sqrt{A(r)B(r)}}{\sqrt{A(r) C(r)+P^2(r)}}\\
&\times \left\{ \frac{p_\varphi}{\omega_0}-\frac{P(r)}{A(r)} +\frac{A(r) C(r)+P^2(r)}{\omega_0A(r)}V^\varphi(r) \Big[ (a_0(r)-1) \, (\omega_0 V^0(r)-p_\varphi V^\varphi(r))+\frac{1}{2}a_1(r) \Big] \right\}\nonumber\\
&\times \left\{\frac{A(r)C(r)+P^2(r)}{A^2(r)} \bigg( 1-\frac{A(r)}{\omega^2_0}\mathcal{C}_{r1}(r,p_\varphi) \bigg)-\left(\frac{p_\varphi}{\omega _0}
-\frac{P(r)}{A(r)}\right)^2\right\} ^{-1/2}, \nonumber
\end{align}
\end{widetext}
where $\mathcal{C}_{r1}(r,p_\varphi)$ remains the same as (\ref{Cr1_gen}).

The formula \eqref{integrand_rotating_n} for the orbit equation is not particularly useful for expressing the combination $p_\varphi/\omega_0$ in terms of the closest radial distance $R$ of the light ray to the gravitating object. It is therefore convenient to introduce new functions, additionally to functions $\mathcal{A}_r$, $\mathcal{B}_r$, $\mathcal{C}_r$ discussed in previous subsections, that serve better for this purpose. They can be defined as
\begin{align}
\mathcal{A}_{\varphi1}(r)&=(1-a_0(r))(V^\varphi(r))^2 \, , \\
\mathcal{B}_{\varphi1}(r)&= \Big[ (a_0(r)-1) \, \omega_0V^0(r)+\frac{1}{2}a_1(r) \Big] V^\varphi(r),\\
\mathcal{C}_{\varphi1}(r)&=\left[ (1-a_0(r)) \, \omega_0 V^0(r)-a_1(r) \right]\omega_0 V^0(r)-a_2(r) \,.
\end{align} 
These expressions do not depend on the metric coefficients and occur solely due to the presence of the medium and its velocity components. Moreover, they are independent of $p_\varphi$, which will be advantageous in the next steps of our analysis. Note that in a static medium, i.e., when $V^\varphi=0$, it holds $\mathcal{A}_{\varphi1}(r)=\mathcal{B}_{\varphi1}(r)=0$.

With the defined relations at hand, it turns out that several expressions in~\eqref{integrand_rotating_n} can be rearranged with the help of these functions as
\begin{align}
V^\varphi(r)& \Big[ (a_0(r)-1) \, (\omega_0 V^0(r)-p_\varphi V^\varphi(r))+\frac{1}{2}a_1(r) \Big]\nonumber\\
&=p_\varphi\mathcal{A}_{\varphi1}(r)+\mathcal{B}_{\varphi1}(r),\\
\mathcal{C}_{r1}(r,p_\varphi)& =p_\varphi^2\mathcal{A}_{\varphi1}(r) +2p_\varphi\mathcal{B}_{\varphi1}(r)+\mathcal{C}_{\varphi1}(r).
\end{align}

The orbit equation (\ref{integrand_rotating_n}) can be finally written as 
\begin{equation}\label{integrand_rotating_compact}
  \frac{d\varphi}{dr} = \pm \sqrt{\mathcal{A}_\varphi(r)B(r)} \,
  \bigg\{ \frac{\mathcal{B}_\varphi^2(r) -\mathcal{A}_\varphi(r)\mathcal{C}_{\varphi}(r)}{\left(\mathcal{A}_\varphi(r)p_\varphi+\mathcal{B}_\varphi(r)\right)^2} - 1 \bigg\}^{-1/2},
\end{equation}
where a new set of functions was utilized, namely
\begin{align}
\mathcal{A}_\varphi(r)&=\mathcal{A}_{\varphi1}(r)+\frac{A(r)}{A(r) C(r)+P^2(r)}\, , \\
\mathcal{B}_\varphi(r)&=\mathcal{B}_{\varphi1}(r)-\frac{\omega_0P(r)}{A(r) C(r)+P^2(r)},\\
\mathcal{C}_\varphi(r)&=\mathcal{C}_{\varphi1}(r)-\frac{\omega^2_0C(r)}{A(r)C(r)+P^2(r)} \,.
\end{align} 
These functions are introduced primarily to compactify the final formula, but they are also useful for isolating contributions of the spacetime metric and medium.
Eq.~(\ref{integrand_rotating_compact}) appears almost identical to the expression (100) derived in \cite{Bezdekova-2024}, but here it is introduced with more general definitions of the functions $\mathcal{A}_\varphi$, $\mathcal{B}_\varphi$, and $\mathcal{C}_\varphi$. The previous result is recovered by setting $P(r)=0$.

The next step in obtaining an analytical formula for the deflection angle is to express the ratio $p_\varphi/\omega_0$ explicitly in terms of the radius of the closest approach $R$. To do so, it is useful to introduce the function
\begin{equation}
\label{eq:hrVphi} h(r)=\frac{1}{\mathcal{A}_\varphi(r)}\sqrt{\mathcal{B}^2_\varphi(r) - \mathcal{A}_\varphi(r) \, \mathcal{C}_\varphi(r)} \, .
\end{equation}
This function should not be confused with $h(r)$ from the previous section; see also Eq.~(102) in Ref.~\cite{Bezdekova-2024}. Using the definitions introduced above, Eq.~\eqref{eq:hrVphi} can be written as
\begin{widetext}
\begin{equation}
  h(r)=\frac{\omega_0}{\mathcal{A}_\varphi(r)}\sqrt{\frac{1+\mathcal{K}_\varphi}{A(r) C(r)+P^2(r)}+\left(\frac{V^\varphi(r)}{\omega_0}\right)^2\left[\frac{1}{4}a_1^2(r)+(1-a_0(r))a_2(r)\right]}\,,
\end{equation}
with
\begin{align}
\mathcal{K}_\varphi &= C(r)(V^\varphi(r))^2(1-a_0(r)) + 2 P(r)V^\varphi(r) \Big[(1-a_0(r)) \, V^0(r)-\frac{1}{2}\omega_0^{-1}a_1(r) \Big]\\
& - A(r)\left[ (1-a_0(r)) \, (V^0(r))^2-\omega_0^{-1}a_1(r)V^0(r) - \omega_0^{-2}a_2(r)) \right].\nonumber
\end{align}
\end{widetext}

At $r=R$ we have $dr/d\varphi=0$, from which we find 
\begin{align}
 p_\varphi=-\frac{\mathcal{B}_\varphi(R)}{\mathcal{A}_\varphi(R)}\pm h(R).
\end{align}
Notice that $\mathcal{A}_\varphi$ does \emph{not} depend on $\omega_0$, while $\mathcal{B}_\varphi$ and $h(r)$ do. This is a general equation showing a relationship between $p_\varphi$ and $h(r)$ which can be applied for arbitrary functions following the structure introduced above. In principle, there should \emph{always} exist suitable forms of these relations, allowing one to get an explicit dependence on $\omega_0$. 

Moreover, since the functions $\mathcal{A}_\varphi(r)$, $\mathcal{B}_\varphi(r)$, $\mathcal{C}_\varphi(r)$ were defined in such a way that they are independent of $p_\varphi$, it is possible to separate the dependence on this constant of motion. This is crucial for finding a relation between $p_\varphi/\omega_0$ and $h(r)$.

Having introduced function $h(r)$, we are able to express the deflection angle for the case of a circular medium motion with the given refractive index (\ref{refr_n}) nicely and compact as
\begin{gather}\label{eq:rotational-defang}
\hat{\alpha} =  2 \int \limits_R^\infty \sqrt{B(r)\mathcal{A}_\varphi(r)} \\
\times \Bigg\{ \frac{h^2(r)}{\big( \frac{\mathcal{B}_\varphi(r)}{\mathcal{A}_\varphi(r)} -\frac{\mathcal{B}_\varphi(R)}{\mathcal{A}_\varphi(R)}\pm h(R) \big)^2} - 1 \Bigg\}^{-1/2} dr \, - \, \pi \, .\nonumber
\end{gather}
This expression appears identical to Eq.~(109) derived in \cite{Bezdekova-2024}, except that the precise radial dependence of the functions involved differs.

\section{Slowly moving medium in slowly rotating spacetimes}\label{sec:slowmedium}

In Sections~\ref{sec:synge} and~\ref{sec:AxDefAng}, we studied the propagation of light in a moving medium within axially symmetric spacetimes: first using a general form of the refractive index $n(r,\omega)$, and then the specific form given by Eq.~\eqref{refr_n}.

An interesting discussion that arises in axially symmetric spacetimes is how the rotational effects of the spacetime coexist with the motion of the medium. In particular, if the medium is rotating around the central object, is it possible for the effects of spacetime rotation and medium rotation to cancel each other under certain conditions?

Within the framework developed in the previous sections, it is difficult to answer this question quantitatively (especially analytically) due to the complexity of the resulting equations. Even though we used a specific form of the refractive index to obtain a closed-form expression for the deflection angle, the resulting equations still contain a mixture of terms of different origins.

To explore the superposition of central object rotation and medium motion more transparently, we now introduce the approximation of a slowly moving medium in this section and apply the perturbative approach. (Not to be confused with the conceptually very different perturbations studied in \cite{Bezdekova-2024}, where perturbations of the refractive index around a cold plasma have been considered.)

To do so, we begin in Subsection~\ref{ssec:slowmedium-A} by taking the general Hamiltonian \eqref{Hamiltonian-1} and assuming that the spatial components of the medium’s 4-velocity are small in the coordinate system under consideration. 
Then, in Subsection~\ref{ssec:slowmedium-B} we focus on axially symmetric spacetimes and derive the expression for the orbit equation $\dot\varphi/\dot r$ to the first order in the medium's velocities, in the equatorial plane. Finally, in Subsection~\ref{ssec:slowmedium-C}, we make an additional assumption: that the metric component $P = g_{0\varphi}$ is of the same order as the spatial velocity of the medium. This enables us to identify how axially symmetric spacetimes with slow rotation parameter and slowly moving media superpose
with each other in their influence on light deflection. In particular, we show that under very specific conditions, the rotational effects of the spacetime and those of the moving medium can compensate each other.

In Section \ref{sec:Kerr} we will apply our general findings to Kerr spacetimes.

\subsection{The Synge Hamiltonian to linear order in the medium velocity on arbitrary spacetime} \label{ssec:slowmedium-A}

In this subsection, we consider the general form of the Hamiltonian and perform an expansion with respect to small medium velocities, keeping terms up to linear order. This is the only part of the paper where we present formulas without imposing any restrictions on the spacetime geometry and the refractive index of the medium. For this reason, we repeat the general form of certain equations in this subsection for the reader’s convenience.
In the following subsections, as well as throughout the rest of the paper, we restrict our analysis to the equatorial plane of the axisymmetric stationary metric and to medium velocities that depend solely on the radial coordinate.

We start again from the Hamiltonian \eqref{Hamiltonian-1} that encodes the propagation of light inside a medium with refractive index $n(x,\omega)$,
\begin{align}\label{eq:HamGen}
&\mathcal{H}(x,p) \nonumber\\
&= \frac{1}{2} \left\{ g^{ik} p_i p_k - [n^2(x, \omega(x,p)) -1] \omega(x,p)^2 \right\}  \, ,
\end{align}
where the photon frequency is given by $\omega(x,p) = - p_j V^j = - p_0 V^0 - p_\alpha V^\alpha$ and $V= V^0\partial_t + V^\alpha \partial_\alpha$ is the 4-velocity of the medium in some given coordinate system.

In this subsection, we assume that the components $V^i$ of the 4-velocity may, in general, depend on all spacetime coordinates, i.e., $V^i = V^i(x)$.

For the following perturbative analysis, we assume that the spatial components $V^\alpha$ of the 4-velocity $V$ are small compared to unity and can thus be treated as perturbative parameters. All relevant expressions will be expanded to the first order in these spatial velocity components. Note that the condition of smallness of the spatial components of the velocity is fulfilled in a particular coordinate system in which 4-velocity is written, and may not hold for the same physical system when the reference frames are changed.

To begin with, we use the normalization condition~(\ref{eq:norm}) to express the $V^0$ component in terms of the $V^\alpha$ components. As a result of the expansion, we obtain:
\begin{widetext}
\begin{align}
   V^0(x, V^\alpha) 
   &= \frac{1}{g_{00}} \left( - g_{0\alpha} V^\alpha - \sqrt{g_{0\alpha}g_{0\beta}V^\alpha V^\beta - g_{00}(1+ g_{\alpha\beta}V^\alpha V^\beta)} \right)\\
   &=  \frac{1}{\sqrt{-g_{00}}} - \frac{g_{0\alpha}V^\alpha}{g_{00}} + \mathcal{O}(V^2)\,.\nonumber
\end{align}
\end{widetext}
Note that only one combination of signs was left here, in accordance with the discussion below Eq.~\eqref{eq:V0gen}. We also recall that the metric component $g_{00}$ is negative, corresponding to the signature convention $\{-,+,+,+\}$.

With this relation we can interpret the Hamiltonian \eqref{eq:HamGen} and the photon frequency as function of spacetime points $x^i$, the momenta of the light ray $p_i$ as well as of the spatial velocity components $V^\alpha$, i.e.,\
\begin{align}
    \mathcal{H}(x,p) &=\mathcal{H}(x,p, V^\alpha)\,,\\
    \omega(x,p) &= \omega(x,p, V^\alpha)\,.
\end{align}
Since we assume that the $V^\alpha$ are small, we can perform a power series expansion in $V^\alpha$ of the general Hamiltonian -- applicable to any slowly moving medium with an arbitrary refractive index in an arbitrary curved spacetime -- and study the consequences within this general approach. Later in Section \ref{ssec:slowmedium-B}, we specialize to an axially symmetric spacetime in order to derive an explicit formula for the deflection angle of light.

We proceed with the linearization of the Hamiltonian \eqref{eq:HamGen} in three steps:
\begin{itemize}
    \item Step 1 -- Expanding the photon frequency:
    \begin{align}
        &\omega(x,p, V^\alpha)\\ 
        &= - p_j V^j 
        = - p_0 V^0 - p_\alpha V^\alpha\nonumber\\
        &= - \frac{p_0}{\sqrt{-g_{00}}} + \bigg( \frac{g_{0\alpha}p_0}{g_{00}} - p_\alpha \bigg) V^\alpha 
        + \mathcal{O}(V^2)\,.\nonumber
    \end{align}
    From this equation we define the zeroth order photon frequency $\bar\omega$, which corresponds to the one for a medium at rest:
    \begin{align}
        \bar \omega:=\omega(x,p, 0)  =- \frac{p_0}{\sqrt{-g_{00}}}\,.
    \end{align}
    \item Step 2 -- Expanding the refractive index:
    \begin{align}
        &n(x, \omega(x,p, V^\alpha))\\
        &= n(x, \bar \omega) 
        + \left(\frac{\partial n}{\partial \omega}(x,\bar\omega) \right)  \left[
        \left( \frac{g_{0\alpha}p_0}{g_{00}} - p_\alpha \right)  V^\alpha \right]\nonumber\\
       & + \mathcal{O}(V^2)\nonumber
    \end{align}
    \item Step 3 -- Expanding the Hamiltonian:
    \begin{align}\label{eq:HamiltonLinear}
        &\mathcal{H}(x,p, V^\alpha)\\
        &= \frac{1}{2} \left\{ g^{ab}(x)p_ap_b - \left[n(x,\omega)^2-1\right] (p_a V^a)^2 \right\} \nonumber\\
        &= \frac{1}{2} \Bigl\{ g^{ab}(x)p_ap_b + \frac{p_0^2}{g_{00}}(n(x,\bar \omega)^2-1)\nonumber\\ 
        &+ \frac{2 p_0 }{g_{00}^2} 
    \left[ p_0\ n(x,\bar\omega) \left(\tfrac{\partial n}{\partial \omega}(x,\bar\omega) \right) -  \sqrt{-g_{00}} \,(n(x,\bar\omega)^2  - 1) \right] \nonumber\\
    &\times
    (g_{00}p_\alpha - g_{0\alpha} p_0)V^\alpha \Bigr\} + \mathcal{O}(V^2)\,.\nonumber
    \end{align}
\end{itemize}
The Hamiltonian \eqref{eq:HamiltonLinear} describes the propagation of light through a slowly moving medium with arbitrary refractive index, on generally curved spacetime to linear order in the small medium velocity.

In the following, we consider this Hamiltonian in an axially symmetric spacetime with a medium that exhibits the velocity profile of the form 
\begin{align}\label{eq:velprof}
    V^\alpha = (V^r(r,\vartheta),V^\vartheta(r,\vartheta),V^\varphi(r,\vartheta))\,.
\end{align} 
We restrict our analysis to the equatorial plane and derive the corresponding deflection angle.

\subsection{Orbit equation in the equatorial plane} \label{ssec:slowmedium-B}

In axially symmetric spacetimes with a metric defined by \eqref{metric_def4D} and its inverse \eqref{metric_def4Dinv}, we have $g_{00} = -A(r,\vartheta)$, and the only non-vanishing off-diagonal component is $g_{0\varphi} = P(r,\vartheta)$. We employ the same assumptions, including asymptotical flatness, as in the non-perturbative treatment (outlined in Section \ref{sec:synge} below Eq. \eqref{metric_def4Dinv}), in order to be able to discuss the orbit equation in the equatorial plane ($p_\vartheta=0, \, \vartheta=\pi/2$). Under these assumptions, and introducing the photon frequency at infinity $\omega_0$, see \eqref{eq:p0w0}, the Hamiltonian \eqref{eq:HamiltonLinear} can be conveniently rewritten as a quadratic polynomial in $p_r$, i.e.,

\begin{align}\label{eq:HamAxLin}
    \mathcal{H}(x,p, V^\alpha)
    = \, & \frac{1}{2} \left[ \mathcal{A}_r(r) \, p_r^2 + 2\mathcal{B}_r(r) \, p_r + \mathcal{C}_r(r) \right] \nonumber\\
    & + \mathcal{O}(V^2)\, ,
\end{align}
where the coefficients
\begin{align}
    \mathcal{A}_r(r)&=\frac{1}{B(r)} \, , \label{coeff_A}\\
    \mathcal{B}_r(r)
    &= \frac{\omega_0 V^r(r)}{A(r)} \tilde{n}(r,\bar\omega) 
    \equiv \mathcal{B}_{r1}(r) V^r(r)\, , \label{coeff_B} \\
    \mathcal{C}_r(r)
    &=\mathcal{C}_{r0}(r) + \mathcal{C}_{rn}(r)
    +\mathcal{C}_{r1}(r) V^\varphi(r),\label{coeff_C}
\end{align}
with
\begin{align}
    \mathcal{C}_{r0}(r)&=\frac{-\omega_0^2 C(r)-2 \omega_0 p_\varphi P(r)+A(r) p_\varphi^2}{A(r) C(r)+P^2(r)} \,,\\
    \mathcal{C}_{rn}(r)&=-\frac{\omega_0^2}{A(r)}(n(r,\bar \omega)^2-1)\,,\\
    \mathcal{C}_{r1}(r)&=\frac{2 \omega_0 }{A(r)^2} 
    \tilde{n}(r,\bar\omega) (A(r) p_\varphi - P(r) \omega_0 ) \,,
\end{align}
are independent of $p_r$. For the quantities of zeroth order (i.e., $V^\alpha=0$, a medium at rest) we introduced the following notation: the plasma frequency is denoted by
\begin{align} \label{eq:bar-omega-def}
    \bar \omega =\frac{\omega_0}{\sqrt{A(r)}}\,,
\end{align}
and the terms that couple to the zeroth order refractive index $n(r,\bar\omega)$ to the velocity from now on we use the abbreviation
\begin{align}\label{eq:tilden}
  &\tilde{n}(r,\bar\omega)\\
  &=-\omega_0\ n(r,\bar\omega)\left(\tfrac{\partial n}{\partial \omega}(r,\bar\omega) \right) 
  -  \sqrt{A(r)}\ (n(r,\bar\omega)^2   - 1)\,.\nonumber
\end{align}
Observe that for a cold plasma we have that $n(r,\bar\omega)^2 = 1 - \omega_p^2/\bar\omega^2 \frac{}{}$ and thus $\tilde n(r,\bar\omega) = 0$. Moreover, it is useful to define function $\mathcal{C}_{r0}(r,p_\varphi)$ because it directly corresponds to the vacuum case for which $V^\alpha = 0$ and $n(r,\bar\omega) = 0$.

The form of the Hamiltonian \eqref{eq:HamAxLin} is suitable for deriving the orbit equation, $\dot{\varphi}/\dot{r}$, to the first order in the spatial components of $V$. All subsequent expressions should be understood in this perturbative context. The Hamilton equations of motion then yield:
\begin{align}
    \dot \varphi 
    &= \frac{\partial \mathcal{H}}{\partial p_\varphi}\label{eq:dphi}\\
    &=\frac{ -\omega_0 P(r)+A(r) p_\varphi}{A(r) C(r)+P^2(r)}+ \frac{\omega_0 V^\varphi(r)}{A(r)}\tilde{n}(r,\bar\omega),\nonumber\\
    \dot r 
    &= \frac{\partial \mathcal{H}}{\partial p_r} \label{eq:dr}\\
    &=\frac{p_r}{B(r)} + \frac{\omega_0V^r(r)}{A(r)}\tilde{n}(r,\bar\omega)\,,\nonumber
\end{align}
and hence, we find to the first order in the velocity
\begin{align}\label{eq:defang}
    \frac{\dot \varphi}{\dot r}
    &= \frac{B(r)}{p_r} \left(\frac{-\omega_0 P(r)+A(r) p_\varphi}{A(r) C(r)+P^2(r)}+ \frac{\omega_0 V^\varphi(r)}{A(r)}\tilde{n}(r,\bar\omega)\right)\nonumber\\
    &- \tilde{n}(r,\bar\omega) \frac{B(r)^2}{p_r^2}
    \frac{(-\omega_0 P(r)+A(r) p_\varphi)}{(A(r) C(r)+P^2(r))} \frac{\omega_0V^r(r)}{A(r)}\,.
\end{align}
The right-hand side of the orbit equation \eqref{eq:defang} contains the variable $p_r$. To analyze this formula further, we must therefore use the dispersion relation $\mathcal{H} = 0$, which in this section is used in the form of Eq.~\eqref{eq:HamAxLin}. This relation can be applied in several ways. On one hand, we can solve Eq.~\eqref{eq:HamAxLin} directly for $p_r$ and then expand the resulting expression in the small spatial components of the velocity:
\begin{align}
    p_r 
    &= -\frac{-\mathcal{B}_r \pm \sqrt{\mathcal{B}_r^2 - \mathcal{A}_r \mathcal{C}_r }}{\mathcal{A}_r}\\
    &= \mp \frac{ \sqrt{- \mathcal{C}_{r0} - \mathcal{C}_{rn}} }{\sqrt{\mathcal{A}_r}} + \frac{\mathcal{B}_{r1}V^r \pm \tfrac{\sqrt{\mathcal{A}_r}\mathcal{C}_{r0} V^\varphi}{2\sqrt{- \mathcal{C}_{r0} - \mathcal{C}_{rn}}}}{\mathcal{A}_r} \nonumber \, .
\end{align}
On the other hand, one can follow a different approach, outlined in Appendix~\ref{app_no_pr}, which avoids the explicit expression of $p_r$ in terms of $V^\alpha$.

In addition, at the point of the closest approach of the light ray to the gravitating object, $R$, we have that $\dot r = \partial \mathcal{H}/ \partial p_r = 0$, which implies from \eqref{eq:dr} that $p_r(R) = -\omega_0 V^r(R) \tilde n(R,\bar\omega) B(R)/ A(R)$. Using this in $\mathcal{H}(R,p_r(R)) = 0$, we can determine the ratio between the constants of motion $\omega_0$ and $p_\varphi$ as
\begin{align}\label{b_CP}
    \frac{p_{\varphi}}{\omega_{0}} 
    &= \mp \, h(R)+ \frac{P(R)}{A(R)} - \frac{\tilde n(R,\bar\omega) h(R)^2 V^{\varphi}(R)}{n(R,\bar\omega)^2}\,,
\end{align}
with 
\begin{equation}\label{def_h2}
h^2(r)=n(r,\bar \omega)^2\left(\frac{C(r)}{A(r)}+\frac{P^2(r)}{A^2(r)}\right).
\end{equation}
Using these findings in equation \eqref{eq:defang} we can express the orbit equation $\dot \varphi/\dot r = d\varphi/dr$ through the constant $R$ instead of $p_\varphi$:
\begin{widetext}
\begin{align}\label{eq:defint}
\frac{d \varphi}{d r} \, = \, & \pm \, \sqrt{\frac{A(r)B(r)}{A(r) C(r)+P^2(r)}} \, \Biggl\{ \frac{h^2(r)}{ \big( \frac{P(r)}{A(r)} -\frac{P(R)}{A(R)}\pm h(R) \big)^2} - 1 \Biggr\}^{-1/2}\\
& \times  \Bigg\{ 1 + \frac{h^2(r)}{n(R,\bar\omega)^2n(r,\bar\omega)^2} \frac{\left(n(r,\bar\omega)^2 V^\varphi(R) \, \tilde{n}(R,\bar\omega) h^2(R) - n(R,\bar\omega)^2 V^\varphi(r) \tilde{n}(r,\bar\omega) h^2(r) \right) }{\big( \frac{P(r)}{A(r)}-\frac{P(R)}{A(R)}\pm h(R) \big) \Big[ h^2(r)-\big( \frac{P(r)}{A(r)}-\frac{P(R)}{A(R)}\pm h(R) \big)^2 \Big] } \Bigg\} \, , \nonumber
\end{align}
where we quickly recall that the functions $A(r)$, $B(r)$, $C(r)$, and $P(r)$ belong to the metric \eqref{metric_def}; $n(r,\omega)$ denotes the refractive index in its general form; $\tilde{n}(r,\bar\omega)$ is given in Eq.~\eqref{eq:tilden}; $\bar\omega$ is defined in Eq.~\eqref{eq:bar-omega-def}; and $h(r)$ is introduced in Eq.~\eqref{def_h2}; the constant $R$ represents the distance of the closest approach; the plus-minus sign must be chosen according to the signs of $d\varphi$ and $dr$ along the corresponding segment of the trajectory.

Remarkably, we find that to the first order in the medium velocity, the radial component $V^r$ of the medium's velocity does not contribute to light deflection in the equatorial plane of an axially symmetric spacetime. The corresponding equation for the orbit equation in particular case of a spherically symmetric spacetime is easily obtained by setting $P(r) = 0$:
\begin{gather}
\frac{d \varphi}{d r} \, = \, \pm \, \sqrt{\frac{B(r)}{C(r)}} \left( \frac{h_0^2(r)}{ h_0^2(R)} - 1 \right)^{-1/2} \left\{ 1 + \frac{h_0^2(r)}{h_0(R)\left (h_0^2(R)- h_0^2(r)\right)}
\biggl( V^\varphi(r)\tilde{n}(r,\bar\omega)\frac{C(r)}{A(r)} -V^\varphi(R)\tilde{n}(R,\bar\omega)\frac{C(R)}{A(R)} \biggr) \right\} \, ,
\end{gather}
\end{widetext}
where
\begin{equation}\label{eq:h0}
  h_0^2(r)  \equiv h^2(r)|_{P=0} = n^2(r,\bar\omega)\frac{C(r)}{A(r)} 
\end{equation}
is the function which has originally been introduced in \cite{Tsupko-2021}, see Eq. (22) there.

Next we will consider the situation in which the deviation of the spacetime geometry from spherical symmetry is small. In particular, we will study the case where this deviation is of the same small order as the velocity of the medium. Then, before we conclude, we will analyze light deflection by a slowly moving medium in a slowly rotating Kerr spacetime in Section~\ref{sec:Kerr}.

\subsection{Slowly rotating gravitating objects -- linearization in $P(r)$}
\label{ssec:slowmedium-C}

In the previous subsection, we linearized the equations under the assumption of a slowly moving medium. To make the superposition of the rotational effects of the spacetime and the medium more transparent, we now consider the case of a slowly rotating gravitating object, while maintaining the approximation of a slowly moving medium. This means that we linearize the orbit equation \eqref{eq:defint} further, by expanding it to the first order in $P(r)$.
In this approximation, we treat $P(r)$ as being of the same order of smallness as the azimuthal component $V^\varphi(r)$ of velocity. (Recall that we have shown the radial component $V^r$ does not contribute at linear order.) Consequently, we neglect higher-order mixed terms proportional to $V^\varphi(r) P(r)$.

To perform the linearization in $P$, we first examine how this metric component manifests in \eqref{eq:defint}. The right-hand side of \eqref{eq:defint} has the form $F(P)(1+ G(P)V^\varphi)$, where $F(P)$ and $G(P)$ are complicated functions. Here we omit explicit dependencies on other metric functions for clarity. Expanding this expression to the leading order in both $P$ and $V^\varphi$ yields:
\begin{align}\label{eq:Plin1}
    &F(P)(1+ G(P)V^\varphi)\\
    &\approx \Bigl( F(0)+  \frac{dF}{dP} \Big|_{P=0} P \Bigr)\left(1+ G(0)V^\varphi\right)\nonumber\\
    &\approx F(0) \Bigl( 1 + \frac{dF}{dP} \Big|_{P=0} \frac{P}{F(0)} + G(0)V^\varphi \Bigr) \, , \nonumber
\end{align}
which corresponds to the first order in $P$.

\begin{widetext}
Explicitly, the function $F(P)$ and its leading order reads
\begin{align}\label{eq:F(P)lin}
    F(P) 
    & \, = \, \pm \, \sqrt{\frac{A(r) B(r)}{A(r) C(r)+P(r)^2}} \, \Bigg\{ \frac{h(r)^2}{\big( \frac{P(r)}{A(r)} - \frac{P(R)}{A(R)} \pm h(R) \big)^2} - 1 \Bigg\}^{-1/2} \\
    & \, \approx \, \pm \sqrt{\frac{B(r)}{C(r)}} \left(\frac{h_0(r)^2}{h_{0}(R)^2}-1\right)^{-1/2}
    \bigg\{ 1 \pm \frac{h_0(r)^2}{h_{0}(R)\left(h_{0}(R)^2 - h_0(r)^2\right)}
    \bigg( \frac{P(r)}{A(r)} - \frac{P(R)}{A(R)} \bigg) \bigg\} \nonumber \, ,
\end{align}
where we again use \eqref{eq:h0} as it was already introduced in \cite{Tsupko-2021} for spherically symmetric spacetimes. From \eqref{eq:defint} we can further conclude that the function $G(0)$ is of the form
\begin{equation} \label{eq:G(0)}
    G(0) =  \frac{h_0(r)^2}{h_{0}(R)\left(h_{0}(R)^2- h_0(r)^2\right)} 
    \left( \frac{V^\varphi(r)\tilde{n}(r,\bar\omega) C(r)}{A(r)} - \frac{V^\varphi(R) \tilde{n}(R,\bar\omega) C(R)}{A(R)}   \right) \,.
\end{equation}

Using \eqref{eq:F(P)lin} and \eqref{eq:G(0)} to evaluate \eqref{eq:Plin1}, we can express the orbit equation \eqref{eq:defint} up to linear order with respect to $P$ and $V^\varphi$ as
\begin{align}\label{eq:dphidrlinpv}
    \frac{\dot \varphi}{\dot r} 
    \, = \, & \pm \sqrt{\frac{B(r)}{C(r)}}\left(\frac{h_0(r)^2}{h_{0}(R)^2}-1\right)^{-1/2}\\
    & \times \left\{ 1 + \frac{h_0(r)^2}{h_{0}(R)\left(h_{0}(R)^2- h_0(r)^2\right)} 
    \bigg( \frac{P(r) + V^\varphi(r) \, \tilde{n}(r,\bar\omega) C(r)}{A(r)}  - \frac{P(R) + V^\varphi(R) \, \tilde{n}(R,\bar \omega) C(R)}{A(R)} \bigg) \right\} \,.\nonumber
\end{align}
\end{widetext}
From this expression, we can make the interesting observation about the simultaneous presence of spacetime rotation effects and those induced by the rotation of the medium. If the velocity profile of the medium is given by
\begin{align}\label{eq:rotation-cancel}
    V^\varphi(r) = - \frac{P(r)}{\tilde n(r,\bar\omega) C(r)}\,,
\end{align}
then the first-order effects precisely cancel, and thus the motion of the medium exactly compensates for the combined effects of black hole rotation and refractive index. Conversely, this condition can also be used to determine a specific refractive index as a function of the spacetime geometry and the velocity profile, such that the first-order effects vanish. This result explicitly demonstrates how black hole rotation and medium motion can produce equivalent observational signatures. In particular, in carefully tuned configurations, these effects may completely cancel one another. 

Next we will evaluate the linearized orbit equation \eqref{eq:dphidrlinpv} and the medium/spacetime rotation balance equation~\eqref{eq:rotation-cancel} for slowly rotating Kerr black holes.

\section{Example: Slowly rotating Kerr metric with slowly moving medium}\label{sec:Kerr}

The components of the Kerr metric in the equatorial plane are given by (for details, see for example \cite{Bezdekova-2023} and references therein) 
\begin{align}
  A(r) &= 1-\frac{2m}{r},  &B(r)& = \left(1 -\frac{2m}{r}+ \frac{a^2}{r^2}\right)^{-1}, \nonumber \\
  C(r) &= r^2 + a^2 +\frac{2ma^2}{r}, &P(r)& =- \frac{2ma}{r}\,,
\end{align}
where $m$ is the mass parameter and $a$ the rotation parameter of the black hole. To the first order in the rotation parameter they simplify to
\begin{align}
  A(r) &= 1-\frac{2m}{r},  &B(r)& = \left(1 -\frac{2m}{r}\right)^{-1}, \\
  C(r) &= r^2, &P(r)& =- \frac{2ma}{r}\,.\nonumber
\end{align}

This situation matches the case we discussed in the previous section, when we assume that the velocity of the medium surrounding the black hole is of the same order as the small rotation parameter. Using these expressions in \eqref{eq:dphidrlinpv} we find
\begin{widetext}
\begin{align}
    \frac{\dot \varphi}{\dot r} 
    \, = \, & \pm \frac{\sqrt{B(r)}}{r} \left(\frac{r^2  B(r) n(r,\bar \omega)^2 }{R^2 B(R) n(R,\bar \omega)^2}-1\right)^{-1/2} \
    \bigg\{ 1 + \frac{r^2\ n(r,\bar\omega)^2B(r)}{R\ n(R,\bar\omega)\sqrt{B(R)}}\frac{\left( Q(r)B(r) 
    -  Q(R)B(R)  \right)}
    {\left( R^2 n(R,\bar\omega)^2 B(R) - r^2 n(r,\bar\omega)^2 B(r) \right)} 
    \bigg\} \\
    \, = \, & \pm \frac{1}{\sqrt{r(r-2m)}}\left(\frac{r^3(R-2m)n^2(r,\bar\omega)}{R^3(r-2m)n^2(R,\bar\omega)}-1\right)^{-1/2}\nonumber \\
    & \times \bigg\{ 1 + \frac{r^3(R-2M)^{3/2}n^2(r,\bar\omega)}{R^{3/2}n(R,\bar\omega)\left(R^3(r-2m)n^2(R,\bar\omega)- r^3(R-2m)n^2(r,\bar\omega)\right)}
    \left(\frac{r Q(r)}{r-2m}  - \frac{R Q(R)}{R-2m} \right) \bigg\} \, ,\nonumber
\end{align}
\end{widetext}
where the function $Q(r)$ quantifies the balance between spacetime rotational and  medium rotational effects,
\begin{align}
    Q(r) = r^2 \tilde n(r,\bar\omega) V^\varphi(r)-\frac{2 a m}{r} \, .
\end{align}
We see explicitly that a cancellation of effects occurs when $Q(r)=0$, meaning that
\begin{align}\label{eq:Kerr-norotation}
   \tilde n(r,\bar\omega) V^\varphi(r) =\frac{2 a m}{r^3} \,.
\end{align}
We remind that $\tilde{n}(r,\bar\omega)$ is defined in Eq.~\eqref{eq:tilden}.

\section{Discussion and Conclusions} \label{sec:discussion}

Gravitational deflection and resulting effects of gravitational lensing represent an excellent probe of our understanding of gravity as described by the geometry of spacetime. Observations of lensing phenomena yield information about both the weak and strong gravity regimes. While the former is already supported by an extensive collection of observational evidence, the latter has only recently been demonstrated through the groundbreaking work of the Event Horizon Telescope.

Apart from gravity, the propagation of light near gravitating objects can be influenced by other factors. These include the interaction of electromagnetic waves with matter, even when its gravitational influence is negligible.
In particular, near black holes or other compact objects, the matter density can become significant due to the formation of accretion flows, and its impact must therefore be carefully examined. 
Within the geometrical optics approximation, the presence of a surrounding medium primarily manifests through refraction, leading to the additional light bending. Analytical studies of gravitational lensing in such environments have mainly focused on the case of cold non-magnetized plasma (see Sections \ref{sec:intro} and \ref{sec:previous} for more details).
The standard approach for this analysis, detailed in Section \ref{sec:synge}, employs the Hamiltonian formalism established by Synge \cite{Synge-1960} for describing light ray propagation in a curved spacetime filled with a dispersive medium; see also Perlick \cite{Perlick-2000}.

In this paper, we have considered an axially symmetric (i.e., rotating) spacetime filled with a moving dispersive medium characterized by a refractive index. We have analytically investigated the influence of the medium's motion on the deflection of light propagating near the gravitating object. Our analysis extends beyond the commonly used cold plasma case, for which the motion of the medium  is known to have no influence. Moreover, by simultaneously accounting for the rotation of both the gravitating object and the surrounding medium, we have been able to investigate the interplay between these two rotational effects.

Following the discussion of Synge's approach and the corresponding equations of motion for the general case of a stationary axially symmetric spacetime with an arbitrary refractive index (Section \ref{sec:synge}), we have selected, in Section \ref{sec:AxDefAng}, a specific yet sufficiently general form for the refractive index \eqref{refr_n}. This form has allowed us a fully analytical treatment and has encompassed physically relevant particular cases.

We have derived the orbit equation \eqref{integ_mixed_n} for light deflection in the equatorial plane of a spacetime filled with a medium possessing both radial and azimuthal velocity components (Subsection \ref{ssec:mixedCase}). Subsequently, in Subsections \ref{ssec:radialMotion} and \ref{ssec:rotatingMotion}, we have obtained closed-form expressions for the deflection angle in two special and instructive cases: for a purely radially moving medium in Eq.~\eqref{eq:defl-angle-final} and for a purely rotating medium in Eq.~\eqref{eq:rotational-defang}.
Noteworthy, even in axial symmetry, the analytic expressions in the equatorial plane take a compact form and naturally extend the previous spherical-symmetry results \cite{Bezdekova-2024} to the physically more realistic case of rotating spacetimes.

To quantitatively analyze the simultaneous presence of the rotational effects of spacetime and those induced by the medium's motion on light propagation, we have considered the case of a slowly moving medium (Section \ref{sec:slowmedium}). We have employed a perturbative approach, treating the medium's velocity as a small parameter, and derived the Hamiltonian \eqref{eq:HamiltonLinear} governing the propagation of light in a generic curved spacetime with a slowly moving medium. Then, we used this expression to obtain the orbit equation \eqref{eq:defint} in the equatorial plane of an stationary axially symmetric spacetime, to the first order in the medium velocity. 
Remarkably, without any further assumptions, we have found that the radial component of the medium's velocity does not contribute at this order; only the rotational component enters. Furthermore, if the spacetime itself is also slowly rotating -- i.e., by linearizing in both the medium's velocity and the off diagonal  metric component -- we can quantify the relation between the rotational component of medium's velocity and the rotation parameters of the spacetime such that these effects cancel, see Eq.~\eqref{eq:rotation-cancel}. In the particular case of a Kerr black hole, this cancellation can be directly related to the angular momentum parameter of a rotating black hole, see Eq.~\eqref{eq:Kerr-norotation}. 

With our work, we have demonstrated that there exists a degeneracy between rotational effects arising from spacetime geometry and those induced by the motion of a medium in light deflection. These effects can potentially lead to indistinguishable observational signatures and therefore must be carefully disentangled. In future work, we aim to extend our analysis from light deflection to the calculation of lensing images. In the weak-deflection regime, more compact expressions for the deflection angle are expected, in contrast to the present study where no assumption on the deflection strength is made. In addition, the influence of different choices of refractive indices will be investigated.

\section*{Acknowledgments}

C.P. acknowledges support by the Deutsche Forschungsgemeinschaft (DFG, German Research Foundation) under Germany's Excellence Strategy -- EXC-2123 QuantumFrontiers -- 390837967 and was funded by the Deutsche Forschungsgemeinschaft (DFG, German Research Foundation) -- Project Number 420243324.

B.B. thanks Professor Pavel Krtou\v{s} and the Institute of Theoretical Physics at the Charles University in Prague for supporting her research during autumn 2024, when the core results of this paper were derived.

O.Y.T. thanks the Excellence Cluster QuantumFrontiers of the German Research Foundation for supporting his visit to ZARM, University of Bremen, and Eva Hackmann and her group for their kind hospitality.

The authors thank Volker Perlick for useful discussions.

\appendix

\section{Alternative derivation of the orbit equation (\ref{eq:dphi}) in Section \ref{sec:slowmedium}}\label{app_no_pr}

This appendix provides an alternative derivation of the orbit equation (\ref{eq:dphi}). In Subsection~\ref{ssec:slowmedium-B}, the intermediate steps leading to Eq.~(\ref{eq:dphi}) involved expressing $p_r$ explicitly in terms of the velocity components, expanded to linear order.

One can circumnavigate this step, which might seem cumbersome, by adapting the same procedure as was applied in \cite{Bezdekova-2024} and by working with the radial equation of motion in the form
\begin{align} \label{eq:app-01}
    \dot r 
    &= \partial \mathcal{H} / \partial p_r \\
    &=\mathcal{A}_r(r) \, p_r + \mathcal{B}_r(r) \, ,  \nonumber
\end{align}
which follows from the quadratic structure of the Hamiltonian (\ref{eq:HamAxLin}) in $p_r$.

To find the expression for $p_r$, one solves $\mathcal{H}=0$. As Eq.~(\ref{eq:HamAxLin}) is quadratic in $p_r$, its solutions are
\begin{equation}\label{sol_pr}
    p_r = \frac{- \mathcal{B}_r(r) \pm \sqrt{\mathcal{B}^2_r(r)-\mathcal{A}_r(r)\mathcal{C}_r(r)}}{\mathcal{A}_r(r)}.
\end{equation}
Substituting this expression into the right hand side of Eq.~\eqref{eq:app-01} yields
\begin{equation}
  \dot r= \pm \sqrt{ \mathcal{B}^2_r(r) - \mathcal{A}_r(r) \, \mathcal{C}_r(r) } \, ;
\end{equation}
compare with Eqs.~(45), (46), and (47) of Ref.~\cite{Bezdekova-2024}.

Plugging the expressions (\ref{coeff_A}), (\ref{coeff_B}), (\ref{coeff_C}) into the corresponding places gives
\begin{gather}\label{B2_AC}
\mathcal{B}^2_r(r)-\mathcal{A}_r(r) \, \mathcal{C}_r(r)=\\
\frac{\omega_0^2}{A(r)B(r)} \bigg\{ n(r,\bar \omega)^2-\frac{A^2(r)}{A(r) C(r)+P^2(r)}\left(-\frac{p_\varphi}{\omega_0}+\frac{P(r)}{A(r)}\right)^2 \nonumber\\
+ 2 \, \tilde{n}(r,\bar\omega)V^\varphi(r) \biggl( -\frac{p_\varphi}{\omega_0}  + \frac{P(r)}{A(r)} \bigg) \bigg\} . \nonumber 
\end{gather}

An apparent benefit of this approach is especially useful in the discussed linearization approach. Even from (\ref{B2_AC}) one can see that term $\propto V^r$ drops out in the linearization process because it represents a contribution of higher order.

Moreover, setting expression (\ref{B2_AC}) to equal zero can be used to find the relation between $-p_\varphi/\omega_0$ and $R$. In the linear order this  returns
\begin{align}
-\frac{p_\varphi}{\omega_0}+\frac{P(R)}{A(R)}
\approx
&\pm h(R)\\
&\pm\frac{\tilde{n}(R,\bar\omega)V^\varphi(R)h(R)}{n(R,\bar \omega)^2}\left(-\frac{p_\varphi}{\omega_0}+\frac{P(R)}{A(R)}\right),\nonumber
\end{align}
where the relation for $h(r)$ is given by (\ref{def_h2}).
From that, one can easily express the ratio $-p_\varphi/\omega_0$ which reads
\begin{gather}\label{b_App}
-\frac{p_\varphi}{\omega_0}=\pm h(R)-\frac{P(R)}{A(R)}+\frac{\tilde{n}(R,\bar\omega)V^\varphi(R)h^2(R)}{n(R,\bar \omega)^2}.
\end{gather}

With the help of these relations and Eq.~(\ref{eq:dphi}), the orbit equation to the first order in the velocity is expressed as
\begin{widetext}
\begin{gather}
    \frac{\dot \varphi}{\dot r}=\pm\left(\frac{A(r) p_\varphi-\omega_0 P(r)}{A(r) C(r)+P^2(r)}+ \frac{\omega_0 V^\varphi(r)}{A(r)}\tilde{n}(r,\bar\omega)\right)\left(\mathcal{B}^2_r(r)-\mathcal{A}_r(r)\mathcal{C}_r(r,p_\varphi)\right)^{-1/2} \\
    \approx \pm \left(\frac{ A(r) p_\varphi-\omega_0 P(r)}{A(r) C(r)+P^2(r)}+ \frac{\omega_0 V^\varphi(r)}{A(r)}\tilde{n}(r,\bar\omega)\right) \bigg\{ \frac{\omega_0^2}{A(r)B(r)} \biggl[ n(r,\bar \omega)^2-\frac{A^2(r)}{A(r) C(r)+P^2(r)} \biggl( -\frac{p_\varphi}{\omega_0}+\frac{P(r)}{A(r)} \biggr)^2 \nonumber\\
 + \, 2 \, \tilde{n}(r,\bar\omega)V^\varphi(r) \biggl( -\frac{p_\varphi}{\omega_0}  + \frac{P(r)}{A(r)} \biggr) \biggr] \bigg\}^{-1/2},\nonumber
\end{gather}
which after some algebraic manipulations and by applying (\ref{b_App}) returns
\begin{align}
\frac{\dot \varphi}{\dot r} \, = \, & \pm \, \sqrt{\frac{A(r)B(r)}{A(r) C(r)+P^2(r)}} \, \Biggl\{ \frac{h^2(r)}{ \big( \frac{P(r)}{A(r)} -\frac{P(R)}{A(R)}\pm h(R) \big)^2} - 1 \Biggr\}^{-1/2}\\
& \times  \Bigg\{ 1 + \frac{h^2(r)}{n(R,\bar\omega)^2n(r,\bar\omega)^2} \frac{\left(n(r,\bar\omega)^2 V^\varphi(R) \, \tilde{n}(R,\bar\omega) h^2(R) - n(R,\bar\omega)^2 V^\varphi(r) \tilde{n}(r,\bar\omega) h^2(r) \right) }{\big( \frac{P(r)}{A(r)}-\frac{P(R)}{A(R)}\pm h(R) \big) \Big[ h^2(r)-\big( \frac{P(r)}{A(r)}-\frac{P(R)}{A(R)}\pm h(R) \big)^2 \Big] } \Bigg\} \, , \nonumber
\end{align}
The obtained expression is identical to formula~(\ref{eq:defint}).
\end{widetext}

\bibliography{MovingMedium_1/biblio}

\end{document}